\documentclass[a4paper,11pt]{article}
\pdfoutput=1 
\usepackage{jheppub} 
\usepackage[T1]{fontenc} 
\usepackage{multirow}
\usepackage[normalem]{ulem}
\usepackage{graphicx}
\usepackage{amsmath,amssymb}
\usepackage{dcolumn}
\usepackage{bm}
\usepackage{color}

\newcommand{\f}{\frac}
\newcommand{\lt}{\left}
\newcommand{\m}{m_{\rm P}}
\newcommand{\n}{\nonumber}
\newcommand{\p}{\partial}
\newcommand{\rt}{\right}
\newcommand{\dd}{{\rm d}}
\newcommand{\bt}{\beta}
\newcommand{\dt}{\delta}

\newcommand{\ve}{\varepsilon}
\newcommand{\sg}{\sigma}

\newcommand{\pb}{{\rm PBH}}
\newcommand{\cR}{{\cal R}}
\newcommand{\cP}{{\cal P}}
\newcommand{\la}{\langle}
\newcommand{\ra}{\rangle}
\newcommand{\arxgr}[1]{\href{http://arxiv.org/abs/#1}{{\ttfamily arXiv:#1[gr-qc]}}}
\newcommand{\arxph}[1]{\href{http://arxiv.org/abs/#1}{{\ttfamily arXiv:#1[hep-ph]}}}
\newcommand{\arxth}[1]{\href{http://arxiv.org/abs/#1}{{\ttfamily arXiv:#1[hep-th]}}}
\newcommand{\arxex}[1]{\href{http://arxiv.org/abs/#1}{{\ttfamily arXiv:#1[hep-ex]}}}
\newcommand{\arxas}[1]{\href{http://arxiv.org/abs/#1}{{\ttfamily arXiv:#1[astro-ph]}}}
\newcommand{\Arxgr}[1]{\href{http://arxiv.org/abs/gr-qc/#1}{{\ttfamily arXiv:#1[gr-qc]}}}
\newcommand{\Arxth}[1]{\href{http://arxiv.org/abs/hep-th/#1}{{\ttfamily arXiv:#1[hep-th]}}}
\newcommand{\Arxas}[1]{\href{http://arxiv.org/abs/astro-ph/#1}{{\ttfamily arXiv:#1[astro-ph]}}}
\newcommand{\Arxph}[1]{\href{http://arxiv.org/abs/hep-ph/#1}{{\ttfamily arXiv:#1[hep-ph]}}}

\title{\boldmath Primordial black holes and scalar-induced gravitational waves from the perturbations on the inflaton potential in peak theory}
\author[a]{Ji-Xiang Zhao}
\author[a]{Xiao-Hui Liu}
\author[a]{Nan Li}
\affiliation[a]{Department of Physics, College of Sciences, Northeastern University \\ No. 3-11, Wenhua Road, Shenyang, 110819, China \\}
\emailAdd{2100189@stu.neu.edu.cn}
\emailAdd{liuxiaohui22@mails.ucas.ac.cn}
\emailAdd{linan@mail.neu.edu.cn}

\abstract{A perturbation on the background inflaton potential can lead inflation into the ultraslow-roll stage and can thus remarkably enhance the power spectrum ${\cal P}_{\cal R}(k)$ of the primordial curvature perturbation on small scales. Such an enhanced ${\cal P}_{\cal R}(k)$ will result in primordial black holes (PBHs), contributing a significant fraction of dark matter, and will simultaneously generate sizable scalar-induced gravitational waves (SIGWs) as a second-order effect. In this work, we calculate the PBH abundances $f_{\rm PBH}(M)$ and SIGW spectra $\Omega_{\rm GW}(f)$ in peak theory. We obtain the PBHs with desirable abundances in one or two typical mass windows at $10^{-17}~M_\odot$, $10^{-13}~M_\odot$, and $30~M_\odot$, respectively. At the same time, the relevant SIGWs are expected to be observed by the next-generation gravitational wave detectors, without spoiling the current constraint. Especially, the SIGW associated with the PBH of $30~M_\odot$ can also interpret the potential isotropic stochastic gravitational wave background from the NANOGrav 12.5-year dataset.}

\begin{document}
\maketitle
\flushbottom

\section{Introduction} \label{sec:intro}

The detection of gravitational waves (GWs) from the merger of binary black holes revealed the dawn of the era of multi-messenger astronomy \cite{LIGO}. The GWs propagate almost freely in the Universe once produced, carrying a wealth of information, and thus provide a powerful tool to explore the early Universe. Meanwhile, there are also other possible sources of GWs, such as phase transitions \cite{source1, source2}, reheating after inflation \cite{source4, source5, source6}, topological defects \cite{source8, source9}, etc. Furthermore, the GWs from different sources are uncorrelated and thus generate a stochastic GW background together. The detection of such a stochastic GW background will give us important insight into astrophysics and cosmology, so various GW detectors have been designed with different sensitive frequencies \cite{tance1, tance2, tance3, tance4, detect1, detect2, detect4, detect5, detect6, detect7}. In recent years, the scalar-induced GWs (SIGWs) are receiving increasing research interest, with the source being the second-order effect from the first-order scalar perturbations generated during cosmic inflation \cite{yuan1, Saito:2008jc}. More importantly, if these scalar perturbations are large enough on small scales, they will also produce abundant primordial black holes (PBHs) simultaneously, which can form binary black holes, be the seeds of the supermassive black holes in the galactic centers, and behave as a promising candidate of dark matter (DM) \cite{pbh1, seed1}.

In the radiation-dominated (RD) era of the early Universe, if the density contrast of the radiation field is sufficiently large at the horizon reentry, the over-dense region can collapse to PBHs. Because of the Hawking radiation, the PBHs with mass $M<5\times 10^{-19}~M_\odot$ have already evaporated, and the PBHs with mass $M>5\times 10^{-19}~M_\odot$ can still stably exist today \cite{hjfushe}. The PBH abundance $f_\pb$ is defined as its proportion in DM at present. If $f_\pb\sim 0.1$, the PBHs can be considered as an effective candidate of DM; if $f_\pb\ll 10^{-3}$, its possibility as DM can be safely excluded from the relevant mass range. Albeit various experiments have constrained the upper bounds of $f_\pb$ strictly in different mass ranges, there remains an open mass window at $10^{-17}$--$10^{-13}~M_\odot$, where PBHs are possible to compose all DM (with the lower bound $10^{-17}~M_\odot$ in the asteroid mass range and the upper bound $10^{-13}~M_\odot$ in the sub-lunar mass range) \cite{dm}. Moreover, although it has been confirmed that the PBH in the intermediate mass range ($10$--$10^3~M_\odot$) cannot contribute a significant fraction of DM \cite{dm}, its relevant SIGW is still of cosmological interest. Therefore, in this paper, we will consider the PBHs in the three mass windows at $10^{-17}~M_\odot$, $10^{-13}~M_\odot$, and $30~M_\odot$, respectively.

In cosmological perturbation theory, the scalar and tensor perturbations are decoupled at first order, and there is no source term in the equation of motion for the tensor perturbations. However, the scalar perturbations can alter the quadrupole moment of the radiation field, acting as the source of the second-order tensor perturbations, and thus generate SIGWs in the RD era \cite{ylfs}. Therefore, the SIGWs are present inevitably, accompanying the possible formation of PBHs on small scales and providing a powerful tool to constrain the PBH abundance. Usually, a single-field slow-roll (SR) inflation model leads to a nearly scale-invariant power spectrum of the scalar perturbations (around $10^{-9}$), which has been confirmed by the measurement of the cosmic microwave background (CMB) anisotropies on large scales ($<1~{\rm Mpc^{-1}}$) \cite{CMB}. However, such a power spectrum cannot result in a large enough PBH abundance, and the corresponding SIGWs are so weak and are thus negligible compared with the first-order GWs. Nevertheless, if the SR conditions are violated on small scales ($10^5$--$10^{15}~{\rm Mpc^{-1}}$), which can be realized in the so-called ultraslow-roll (USR) stage in inflation, the situation will become rather different. During this stage, the power spectrum of the scalar perturbations can be significantly enhanced up to $10^{-2}$ on small scales, generating the PBHs with desirable masses and abundances. At the same time, the SIGWs are also amplified and can be sizable or even larger than the first-order GWs \cite{sigw2, sigw4}. Consequently, the question reduces to the design of the specific inflation models with the USR stage, which can increase the power spectrum dramatically on small scales, without spoiling the CMB constraints on large scales.

There are many ways to realize the USR conditions, such as inflection-point inflation \cite{inflection1, inflection2}, critical Higgs inflation \cite{Higgs}, non-minimal coupling $R^2$ inflation \cite{R2}, Higgs--$R^2$ inflation \cite{HR}, etc. In this paper, we consider the situation with one or two perturbations $\dt V$ on the background inflaton potential $V_{\rm b}$. By this means, inflation can be studied on small and large scales separately, without the intractable interference in between \cite{Ozsoy:2018flq, Mishra:2019pzq, Ozsoy:2020kat, Zheng:2021vda, Zhang:2021vak}. Previously, $\dt V$ were commonly adopted to be symmetric (e.g., with the Gaussian form), but in Refs. \cite{liuyichen, wangqing}, the authors chose the antisymmetric $\dt V$ [i.e., a linear function times the Gaussian form, see Eq. (\ref{3para}) in Sec. \ref{sec:one} for more detail]. There are several advantages for this choice. First, such a $\dt V$ can be connected to $V_{\rm b}$ very smoothly on both sides of the USR region, so the inflaton can definitely surmount the perturbation, without the worry of eternal inflation \cite{Zheng:2021vda}. Second, there is no modulated oscillation in the power spectrum, naturally avoiding the over-production of tiny PBHs \cite{Pi:2022zxs}. Third, the fine-tuning problem in PBH physics can be greatly relieved \cite{Mishra:2019pzq}. The present work is a succession of Refs. \cite{liuyichen, wangqing}, and we also utilize the antisymmetric $\dt V$. The introduction of such a $\delta V$ can cause a plateau flat enough on $V_{\rm b}$, making the duration of the USR stage sufficiently long. As a result, the PBH abundance can be greatly enhanced, and the relevant SIGW is also expected to be observed by the next-generation GW detectors.

This paper is organized as follows. In Sec. \ref{sec:pbh}, we study the power spectrum of the primordial curvature perturbation and calculate the PBH mass and abundance in peak theory. In Sec. \ref{SIGW}, the SIGW spectrum in the RD era is briefly reviewed. Then, in Secs. \ref{sec:one} and \ref{sec:two}, we study the power spectra $\cP_\cR(k)$, PBH abundances $f_\pb(M)$, and SIGW spectra $\Omega_{\rm GW}(f)$ in the USR inflation models, with one or two perturbations on the background inflaton potential, so that there can be PBHs with masses in one or two typical mass windows at $10^{-17}~M_\odot$, $10^{-13}~M_\odot$, and $30~M_\odot$, with desirable abundances. We conclude in Sec. \ref{sec:con}. We work in the natural system of units and set $c=\hbar=k_{\rm B}=1$.

\section{Power spectrum and PBH abundance} \label{sec:pbh}

In this section, we show the power spectrum ${\cal P}_{\cal R}(k)$ of the primordial curvature perturbation ${\cal R}$, calculate the PBH mass $M$, and discuss the PBH abundance $f_{\rm PBH}(M)$ in peak theory.

\subsection{Basic equations} \label{sec:basic}

We start from the single-field inflation model, with the corresponding action as
\begin{align}
S=\int\dd^4 x\,\sqrt{-g}\lt[\f{m_{\rm P}^2}{2}R-\frac{1}{2}\p_\mu\phi\p^\mu\phi-V(\phi)\rt], \n
\end{align}
where $\phi$ is the inflaton field, $V(\phi)$ is its potential, $R$ is the Ricci scalar, and $m_{\rm P}=1/\sqrt{8\pi G}$ is the reduced Planck mass. The evolution of the inflaton field obeys the Klein--Gordon equation, which can be written as
\begin{align}
\phi_{,NN}+(3-\ve)\phi_{,N}+\frac{1}{H^2}V_{,\phi}=0. \label{kgphi}
\end{align}
Above, the number of $e$-folds $N$ is defined as $\dd N=H(t)\,\dd t=\dd\ln a$, where $t$ is the cosmic time, $a=e^N$ is the scale factor, and $H=\dot{a}/a$ is the Hubble expansion rate. To characterize the motion of the inflaton field, two parameters are introduced for convenience,
\begin{align}
\ve=-\f{\dot H}{H^2}=\f{\phi_{,N}^2}{2\m^2}, \quad
\eta=-\f{\ddot\phi}{H\dot\phi}=\frac{\phi_{,N}^2}{2m_{\rm P}^2}-\frac{\phi_{,NN}}{\phi_{,N}}. \label{SR}
\end{align}
In the usual SR inflation, $\ve$ and $|\eta|$ are much smaller than 1 and are thus called the SR parameters. However, in the USR stage, their values may even approach ${\cal O}(1)$, which have important influences on the PBH abundance and SIGW spectrum. Furthermore, the Friedmann equation can also be expressed as
\begin{align}
H^2=\frac{V}{(3-\ve)m^2_{\rm P}}. \label{Fried}
\end{align}

Now, we move on to the perturbations on the background spacetime. In the conformal Newtonian gauge, the perturbed metric reads
\begin{align}
\dd s^2&=a^2(\tau)\bigg\{-(1+2\Psi)\,\dd \tau^2+\lt[(1-2\Psi)\dt_{ij}+\f{1}{2}h_{ij}\rt]\dd x^i\dd x^j\bigg\}, \label{metric}
\end{align}
where $\tau=\int\dd \tilde{t}/a(\tilde{t})$ is the conformal time, $\Psi$ is the scalar perturbation, and $h_{ij}$ is the tensor perturbation. (Here, we neglect the vector perturbation and anisotropic stress.) A more convenient gauge-invariant scalar perturbation is the primordial curvature perturbation,
\begin{align}
{\cal R}=\Psi+\frac{H}{\dot{\phi}}\dt\phi, \n
\end{align}
and the equation of motion of its Fourier mode ${\cal R}_k$ is the Mukhanov--Sasaki equation \cite{Mukhanov, Sasaki},
\begin{align}
{\cal R}_{k,NN}+(3+\ve-2\eta){\cal R}_{k,N}+\frac{k^2}{H^2e^{2N}}{\cal R}_k=0. \label{MS}
\end{align}

\subsection{Power spectrum} \label{sec:power}

The primordial curvature perturbation ${\cal R}_k$ can be obtained by numerically solving Eqs. (\ref{kgphi})--(\ref{Fried}) and (\ref{MS}). We are more interested in the dimensionless power spectrum ${\cal P}_{\cal R}(k)$, which is defined at the end of inflation as
\begin{align}
\lt.{\cal P}_{\cal R}(k)=\f{k^3}{2\pi^2}|{\cal R}_{k}|^2\rt|_{k\ll aH}.\n
\end{align}
In the RD era, on the comoving slices, the density contrast $\dt$ can be related to ${\cal R}$ at linear order by \cite{Green:2004wb}
\begin{align}
\dt=\f{4}{9}\lt(\f{k}{aH}\rt)^2{\cal R}, \n
\end{align}
so its dimensionless power spectrum $\cP_{\dt}(k)$ is
\begin{align}
\cP_{\dt}(k)=\f{16}{81}\lt(\f{k}{aH}\rt)^4\cP_{\cR}(k). \n
\end{align}

The PBH abundance can be calculated via $\cP_{\dt}(k)$. For this purpose, we first need to smooth the perturbation over some physical scale, usually taken as $R=1/(aH)$, in order to avoid the non-differentiability and the divergence in the large-$k$ limit of the radiation field. This can be realized by introducing a window function $\widetilde{W}(k,R)$ in the Fourier space \cite{window1, window2, window3}. Below, we choose Gaussian window function $\widetilde{W}(k,R)=e^{-k^2R^2/2}$, and the variance of the smoothed density contrast on the scale $R$ is
\begin{align}
\sigma_{\dt}^2(R)=\la\dt^2({\bf x},R)\ra=\int_{0}^{\infty}\f{\dd k}{k}\,\widetilde{W}^2(k,R)\cP_{\dt}(k), \n
\end{align}
where $\la\cdots\ra$ denotes the ensemble average, and we have used the fact $\la\dt({\bf x},R)\ra=0$ for Gaussian random field. The homogeneity and isotropy of the background Universe guarantee that $\sigma_{\dt}^2(R)$ is independent of a special position ${\bf x}$. Furthermore, the $i$-th spectral moment of the smoothed density contrast is defined as
\begin{align}
\sigma_{i}^2(R)=\int_{0}^{\infty}\f{\dd k}{k}\,k^{2i}\widetilde{W}^2(k,R)\cP_{\dt}(k)=\f{16}{81}\int_{0}^{\infty}\f{\dd k}{k}\,k^{2i}\widetilde{W}^2(k,R)(kR)^4\cP_{\cR}(k), \n
\end{align}
where $i=0,1,2, ...$, and $\sigma_0=\sigma_\dt$ naturally.

\subsection{PBH mass and abundance} \label{sec:masss}

Now, we calculate the PBH mass $M$ and its abundance $f_\pb(M)$. In the Carr--Hawking collapse model \cite{carrhkkk}, $M$ is related to the horizon mass at the time of its formation,
\begin{align}
M=\kappa M_{\rm H}=\f{\kappa}{2GH}, \n
\end{align}
where $M_{\rm H}=1/(2GH)$ is the horizon mass, and $\kappa$ is the efficiency of collapse. In the RD era, $H=1/(2t)$, so $M=\kappa{t}/{G}$.

Utilizing the conservation of entropy in the adiabatic cosmic expansion, we obtain \cite{zs}
\begin{align}
\f{M}{M_{\odot}}=1.13\times10^{15}\lt(\f{\kappa}{0.2}\rt)\lt(\f{g_{\ast}}{106.75}\rt)^{-1/6}\lt(\f{k_{\ast}}{k_{\pb}}\rt)^{2}. \label{M}
\end{align}
where $M_{\odot}=1.99\times10^{30}$ kg is the solar mass \cite{changshu}, $g_{\ast}$ is the effective number of relativistic degrees of freedom of energy density, $k_\ast =0.05~{\rm Mpc^{-1}}$ is the CMB pivot scale for the Planck satellite experiment \cite{CMB}, and $k_{\pb}=1/R$ is the wave number of the PBH that exits the horizon. In the RD era, we have $\kappa=0.2$ and $g_{\ast}=106.75$ \cite{Carr:1975qj}. From Eq. (\ref{M}), all spectral moments $\sg_i(R)$ can be reexpressed in terms of the PBH mass as $\sg_i(M)$.

Furthermore, the PBH mass fraction $\bt_{\pb}(M)$ at the time of its formation is defined as
\begin{align}
\lt.\bt_{\pb}(M)=\f{\rho_\pb(M)}{\rho_{\rm R}}\rt|_{\rm formation}, \n
\end{align}
where $\rho_\pb(M)$ and $\rho_{\rm R}$ are the energy densities of PBH and radiation, respectively. The PBH abundance at present is defined as
\begin{align}
\lt.f_\pb(M)=\f{\rho_{\pb}(M)}{\rho_{\rm DM}}\rt|_{\rm today}, \n
\end{align}
where $\rho_{\rm DM}$ is the energy density of DM. Ignoring the evolution of PBHs (e.g., radiation, accretion, and merger), we can finally relate $f_\pb(M)$ to $\bt_\pb(M)$ as \cite{zs}
\begin{align}
f_{\pb}(M)=1.68\times10^{8}\lt(\f{M}{M_\odot}\rt)^{-1/2}\beta_{\pb}(M). \n
\end{align}

\subsection{Peak theory} \label{sec:peak}

The concrete method to calculate the PBH mass fraction $\beta_\pb$ has long been a controversial issue, and different methods usually lead to great difference in the final results \cite{Yoo:2020dkz}. The most general method is peak theory \cite{peak}, with the peak value being the relative density contrast $\nu$, which is defined as $\nu=\dt/\sigma_{\dt}$, and $\nu_{\rm c}=\dt_{\rm c}/\sigma_{\dt}$ is its threshold. The specific value of $\dt_{\rm c}$ depends on the equation of state of the cosmic media and many other ingredients \cite{Niemeyer:1999ak, Musco:2004ak, Musco:2008hv, Musco:2012au, 414, Nakama:2013ica, Musco:2018rwt, Escriva:2019nsa, Escriva:2019phb, Escriva:2020tak, Musco:2020jjb}, and it is the most influential factor in calculating $\beta_\pb$. In this paper, we follow Ref. \cite{414} and set $\dt_{\rm c}=0.414$. However, $\nu_{\rm c}$ is not a constant, as $\sg_\dt$ depends on the smoothing scale $R$.

In peak theory, the number density of peaks is $n({\bf r})=\sum_p\dt_{\rm D}({\bf r}-{\bf r}_p)$, where $\dt_{\rm D}$ is the Dirac function, and ${\bf r}_p$ is the position where the density contrast $\dt$ has a local maximum. This maximum condition needs us to deal with a ten-dimensional joint probability distribution function (PDF) $P(\{y_i\})$ of the Gaussian variables,
\begin{align}
P(\{y_i\})=\f{\exp\big(\f 12\sum_{ij}\Delta y_i{\cal M}^{-1}_{ij}\Delta y_j\big)}{\sqrt{(2\pi)^{10}\det{\cal M}}}, \n
\end{align}
where ${\cal M}$ is the covariance matrix, and $\Delta y_i=y_i-\la y_i\ra$, with $y_1=\dt$, $y_2=\p_1\dt$, ..., $y_5=\p_1\p_1\dt$, ..., and $y_{10}=\p_2\p_3\dt$, respectively. As shown in Ref. \cite{peak}, a series of dimensional reductions can finally reduce the ten-dimensional joint PDF $P(\{y_i\})$ to the one-dimensional conditional PDF $P(\nu)$. By means of $P(\nu)$, the number density of peaks $n(\nu_{\rm c})$ with $\nu>\nu_{\rm c}$ can be written as an integral,
\begin{align}
n(\nu_{\rm c})=\f{1}{(2\pi)^2}\lt(\f{\sg_2}{\sqrt{3}\sigma_{1}}\rt)^3\int_{\nu_{\rm c}}^{\infty}G(\gamma,\nu)e^{-\nu^2/2}\,\dd\nu, \n
\end{align}
where $\gamma=\sigma_{1}^{2}/(\sigma_{\dt}\sigma_{2})$ in the $G(\gamma,\nu)$ function contains the information of the profile of $\dt$. Therefore, the PBH mass fraction $\bt_\pb=n(\nu_{\rm c})V(R)$ can be obtained as
\begin{align}
\beta_\pb=\f{1}{\sqrt{2\pi}}\lt(\f{R\sg_2}{\sqrt{3}\sg_1}\rt)^3\int_{\nu_{\rm c}}^{\infty}G(\gamma,\nu)e^{-\nu^2/2}\,\dd \nu. \n
\end{align}

As the $G(\gamma,\nu)$ function is formally rather complicated, various approximations have been introduced. In Ref. \cite{Green:2004wb}, Green, Liddle, Malik, and Sasaki (GLMS) suggested a very convenient approximation, $\nu\gg1$ and $\gamma\approx1$, meaning that there remain only two independent spectral moments $\sg_\dt$ and $\sg_1$ in $\beta_\pb$. In this approximation, $\beta_\pb$ can be analytically obtained as
\begin{align}
\beta_\pb=\f{1}{\sqrt{2\pi}}\lt(\f{R\sg_1}{\sqrt{3}\sg_\dt}\rt)^3(\nu_{\rm c}^2-1) e^{-\nu_{\rm c}^2/2}. \n
\end{align}
In this paper, we follow the GLMS approximation. For more details about the differences among peak theory, the Press--Schechter theory \cite{PS}, and other approximations of peak theory and their influences on the PBH abundance, see Refs. \cite{p1, p2, p3, wangqing}.

\section{SIGW spectrum} \label{SIGW}

In this section, the SIGW produced in the RD era is reviewed, and the SIGW spectrum at present $\Omega_{\rm GW}(f)$ is also discussed in detail.

\subsection{Basic equations}

First, for the tensor perturbation $h_{ij}(\tau, {\bf x})$ in Eq. (\ref{metric}), its Fourier modes $h_{\bf k}^+(\tau)$ and $h_{\bf k}^\times(\tau)$ are introduced as
\begin{align}
h_{ij}(\tau, {\bf x})=\int\f{\dd^3 k}{(2\pi)^{3/2}}\, e^{i{\bf k\cdot x}}\lt[h_{\bf k}^+(\tau){\rm e}_{ij}^+({\bf k})+h_{\bf k}^\times(\tau){\rm e}_{ij}^\times({\bf k})\rt], \n
\end{align}
where ${\rm e}_{ij}^+({\bf k})$ and ${\rm e}_{ij}^\times({\bf k})$ are two orthonormal polarization tensors. Below, we omit the polarization indices $+$ and $\times$, due to the orthonormal relation $\sum_{i,j}{\rm e}^\alpha_{ij}({\bf k}){\rm e}^\bt_{ij}(-{\bf k})=\dt^{\alpha\bt}$, where $i,j=1,2,3$, and $\alpha,\beta=+,\times$.

The equation of motion of $h_{\bf k}$ can be derived from the perturbed Einstein equations up to second order,
\begin{align}
h''_{\bf k}+2{\cal H}h'_{\bf k}+k^2h_{\bf k}=S(\tau, {\bf k}),\label{eost}
\end{align}
where ${\cal H}=a'/a=aH$ is the comoving Hubble expansion rate, $'$ denotes the derivative with respect to the conformal time $\tau$, and $S(\tau, {\bf k})$ is the Fourier transform of the source term $S_{ij}(\tau, {\bf x})$ \cite{huangqingguo},
\begin{align}
S_{ij}(\tau, {\bf x})&=4\Psi\p_i\p_j\Psi+2\p_i\Psi\p_j\Psi-\f{1}{{\cal H}^2}\p_i(\Psi'+{\cal H}\Psi)\p_j(\Psi'+{\cal H}\Psi).\n
\end{align}
Therefore, the tensor perturbation $h_{ij}$ is induced by the scalar perturbation $\Psi$ as a second-order effect.

Following Ref. \cite{huangqingguo}, we decompose the Fourier mode of $\Psi$ as $\Psi_{\bf k}(\tau)=\Psi(k\tau)\psi_{\bf k}$, where $\psi_{\bf k}$ is the primordial value, and $\Psi(k\tau)$ is the transfer function. In the RD era, we have \cite{main}
\begin{align}
\Psi(k\tau)=\f{9}{(k\tau)^2}\lt[\f{\sin(k\tau/\sqrt{3})}{k\tau/\sqrt{3}}-\cos(k\tau/\sqrt{3})\rt].\n
\end{align}
By this means, the source term $S(\tau, {\bf k})$ in Eq. (\ref{eost}) can be rewritten as
\begin{align}
S(\tau, {\bf k})=\int\f{\dd^3 p}{(2\pi)^{3/2}}\, {\rm e}({\bf k, p})f(\tau, {\bf k, p})\psi_{\bf k}\psi_{{\bf k}-{\bf p}},\label{source}
\end{align}
where ${\rm e}({\bf k, p})={\rm e}^{ij}({\bf k})p_ip_j$ is the projection operator, and $f(\tau, {\bf k, p})$ is the source function,
\begin{align}
f(\tau, {\bf k, p})&=12\Psi(|{\bf p}|\tau)\Psi(|{{\bf k}-{\bf p}}|\tau) +4\tau^2\Psi'(|{\bf p}|\tau)\Psi'(|{{\bf k}-{\bf p}}|\tau) \n\\
&\quad +4\tau\big[\Psi'(|{\bf p}|\tau)\Psi(|{{\bf k}-{\bf p}}|\tau) +\Psi(|{\bf p}|\tau)\Psi'(|{{\bf k}-{\bf p}}|\tau)\big]. \n
\end{align}
Finally, by the Green function method, we can obtain the solution of Eq. (\ref{eost}) in the RD era \cite{Green},
\begin{align}
h_{\bf k}(\tau)=\f{1}{a(\tau)}\int G_k(\tau; \tilde{\tau})a(\tilde{\tau})S(\tilde{\tau}, {\bf k})\,\dd\tilde{\tau},\label{tensor}
\end{align}
where the Green function is $G_k(\tau; \tilde{\tau})={\sin[k(\tau-\tilde{\tau})]}/{k}$.

\subsection{GW spectrum} \label{sec:box}

The GW spectrum $\Omega_{\rm GW}(\tau, k)$ is defined as the GW energy density fraction per logarithmic wave number,
\begin{align}
\Omega_{\rm GW}(\tau, k)=\f{1}{\rho_{\rm c}}\f{\dd\rho_{\rm GW}(\tau, k)}{\dd\ln k}, \label{Omegard}
\end{align}
where $\rho_{\rm GW}$ is the GW energy density, and $\rho_{\rm c}$ is the critical energy density of the Universe. In the transverse--traceless gauge, $\Omega_{\rm GW}(\tau, k)$ can be reexpressed as \cite{ylfs, guanxi2}
\begin{align}
\Omega_{\rm GW}(\tau, k)=\f{1}{24}\lt(\f{k}{\cal H}\rt)^2\overline{{\cal P}_h(\tau, k)}, \label{Omegartt}
\end{align}
where $\overline{(\cdots)}$ denotes the oscillation average, and ${\cal P}_h(\tau, k)$ is the dimensionless power spectrum of the tensor perturbation $h_{\bf k}$,
\begin{align}
{\cal P}_h(\tau, k)=\f{k^3}{2\pi^2}\dt({{\bf k}+{\bf p}})\lt<h_{\bf k}(\tau)h_{\bf p}(\tau)\rt>.\n
\end{align}

From Eqs. (\ref{source}) and (\ref{tensor}), we are able to obtain the two-point correlation function of $h_{\bf k}$,
\begin{align}
\la h_{\bf k}(\tau)h_{\bf p}(\tau)\ra&=\int\f{\dd^3q\dd^3\tilde{q}}{(2\pi)^3}\, {\rm e}({\bf k, q}){\rm e}({\bf p}, \tilde{\bf q})I(\tau,{\bf k, q}) I(\tau, {\bf p}, \tilde{\bf q})\la\psi_{\bf q}\psi_{{\bf k}-{\bf q}}\psi_{\tilde{\bf q}}\psi_{{\bf p}-\tilde{\bf q}}\ra, \label{four}
\end{align}
where $I(\tau, {\bf k, p})$ is the kernel function,
\begin{align}
I(\tau, {\bf k, p})=\int\dd\tilde{\tau}\,\f{a(\tilde{\tau})}{a(\tau)}G_k(\tau; \tilde{\tau})f(\tilde{\tau}, {\bf k, p}). \n
\end{align}
According to the Wick theorem, the four-point correlation function $\la\psi_{\bf q}\psi_{{\bf k}-{\bf q}}\psi_{\tilde{\bf q}}\psi_{{\bf p}-\tilde{\bf q}}\ra$ in Eq. (\ref{four}) can be decomposed into the sum of the products of the two-point correlation functions (or equivalently, the power spectra $\cP_\cR(k)$ of the scalar perturbations) \cite{wick}. For convenience, introducing three dimensionless variables $u=|{\bf k}-{\bf p}|/k$, $v=|{\bf p}|/k$, and $x=k\tau$, we obtain \cite{uv1, uv2}
\begin{align}
{\cal P}_h(\tau, k)&=4\int_{0}^{\infty}\dd v\int_{|1-v|}^{1+v}\dd u\, \lt[\f{4v^2-(1+v^2-u^2)^2}{4uv}\rt]^2{\cal I}^2(x, u, v){\cal P}_{\cal R}(ku){\cal P}_{\cal R}(kv), \label{Ph}
\end{align}
where ${\cal I}(x, u, v)= I(\tau, {\bf k}, {\bf p})k^2$ is the kernel function in terms of the dimensionless variables.

In the RD era, the oscillation average of ${\cal I}^2(x, u, v)$ in the late-time limit of $x\to\infty$ is \cite{main}
\begin{align}
\overline{{\cal I}^2(x\to\infty, u, v)}&=\f12\lt[\f{3(u^2+v^2-3)}{4u^3v^3x}\rt]^2\Bigg\{\lt[(u^2+v^2-3) \ln\lt|\f{(u+v)^2-3}{(u-v)^2-3}\rt|-4uv\rt]^2 \n\\
&\quad+\lt[\pi(u^2+v^2-3)\theta(u+v-\sqrt{3})\rt]^2\Bigg\}, \label{I}
\end{align}
where $\theta$ is the Heaviside step function. It is more convenient to introduce two new variables $t=u+v-1$ and $s=u-v$ for the integral in Eq. (\ref{Ph}). From Eqs. (\ref{Omegard}), (\ref{Ph}), and (\ref{I}), taking into account ${\cal H}=1/\tau$ in the RD era, we finally arrive at \cite{main}
\begin{align}
\Omega_{\rm GW}(\tau, k)&=\f{1}{12}\int_{0}^{\infty}\dd t\,\int_{-1}^{1}\dd s\,\lt[\f{t(t+2)(1-s^2)}{(t+s+1)(t-s+1)}\rt]^2{\cal P}_{\cal R} \lt(\f{t+s+1}{2}k\rt){\cal P}_{\cal R}\lt(\f{t-s+1}{2}k\rt) \n\\
&\quad\times\f{288[t(t+2)+s^2-5]^2}{[(t+s+1)(t-s+1)]^6}\Bigg\{\f{\pi^2}{4}[t(t+2)+s^2-5]^2\theta(t-\sqrt{3}+1) \n\\
&\quad+\lt[\f{1}{2}[t(t+2)+s^2-5]\ln\lt|\f{t(t+2)-2}{3-s^2}\rt|-(t+s+1)(t-s+1)\rt]^2\Bigg\}. \label{Omega}
\end{align}

Although we have obtained the SIGW spectrum $\Omega_{\rm GW}(\tau, k)$ in the early Universe, it will evolve in the cosmic evolution at late times. Well after the horizon reentry, the SIGW produced in the RD era redshifts as radiation, so $\rho_{\rm GW}\varpropto a^{-4}$. Hence, from Eq. (\ref{Omegard}), $\Omega_{\rm GW}$ is constant during the RD era, but is diluted as $a^{-1}$ in the subsequent matter-dominated era. Therefore, the SIGW spectrum at present should be \cite{main}
\begin{align}
\Omega_{\rm GW}(\tau_0, k)=\Omega_{\rm r}^0\Omega_{\rm GW}(\tau_{\rm c}, k),\label{bianhuan}
\end{align}
where $\Omega_{\rm r}^0$ is the current value of the energy density fraction of radiation, and $\tau_{\rm c}$ is some time after $\Omega_{\rm GW}(\tau,k)$ has become constant, so $\Omega_{\rm GW}(\tau_{\rm c},k)$ is an asymptotic constant during the RD era.

To compare with the sensitivity curves of various GW detectors and to understand the relevant physical implications in Secs. \ref{sec:one} and \ref{sec:two}, we emphasize an important relation between the wave number $k$ and the GW frequency $f$ as \cite{Hz}
\begin{align}
f\approx1.5\times10^{-9}\f{k}{{\rm pc^{-1}}}~{\rm Hz}. \label{f}
\end{align}
Combining Eqs. (\ref{Omegartt})--(\ref{f}), we can eventually achieve the present SIGW spectrum $\Omega_{\rm GW}(f)$.

\section{PBHs and SIGWs from one perturbation on the inflaton potential} \label{sec:one}

In this section, we construct one antisymmetric perturbation $\dt V(\phi)$ on the background inflaton potential $V_{\rm b}(\phi)$, in order to achieve the relevant PBH abundances in the GLMS approximation in the three typical mass windows at $10^{-17}~M_\odot$, $10^{-13}~M_\odot$, and $30~M_\odot$, respectively. At the same time, we expect the corresponding SIGWs to be observed by the next-generation GW detectors. Furthermore, we also wish to explain the potential isotropic stochastic GW background from the North American Nanohertz Observatory for Gravitational Waves (NANOGrav) 12.5-year dataset \cite{NANO}. Here, we should stress that we do so mainly to check our inflation model. We do not mean that the NANOGrav signal is the spin-2 SIGW definitely, as it is also consistent with a spin-0 or spin-1 explanation \cite{Sun:2021yra}.

In general, the specific form of $\dt V(\phi)$ is not unique, as long as it can smooth $V_{\rm b}(\phi)$ at some position $\phi_0$. In this way, a plateau appears around $\phi_0$, leading inflation into the USR stage, during which the inflaton field evolves extremely slowly, dramatically enhancing the power spectrum $\cP_\cR(k)$, PBH abundance $f_\pb(M)$, and SIGW spectrum $\Omega_{\rm GW}(f)$ simultaneously. Below, the background inflaton potential $V_{\rm b}(\phi)$ is chosen as the Kachru--Kallosh--Linde--Trivedi potential \cite{KKLT},
\begin{align}
V_{\rm b}(\phi)=V_0\f{\phi^2}{\phi^2+(m_{\rm P}/2)^2}. \n
\end{align}
Furthermore, we follow Refs. \cite{liuyichen, wangqing} and parameterize the antisymmetric perturbation $\dt V(\phi)$ as
\begin{align}
\dt V(\phi)=-A(\phi-\phi_0)\exp\lt[-\f{(\phi-\phi_0)^2}{2\sigma^2}\rt]. \label{3para}
\end{align}
Thus, the inflaton potential reads $V(\phi)=V_{\rm b}(\phi)+\dt V(\phi)$. As $\dt V(\phi)$ is antisymmetric, it can be connected to $V_{\rm b}(\phi)$ very smoothly on both sides of $\phi_0$.

Altogether, there are three parameters in our model: $A$, $\phi_0$, and $\sigma$, characterizing the slope, position, and width of $\dt V(\phi)$, respectively. For convenience, we reparameterize $A$ as
\begin{align}
A=V_{{\rm b},\phi}(\phi_0)(1+A_0), \n
\end{align}
where $A_0$ describes the deviation of $V(\phi)$ from a perfect plateau at $\phi_0$. We set $V_0/m_{\rm P}^4=10^{-10}$, $\phi/\m=3.30$, and $\phi_{,N}/\m=-0.0137$ as the initial conditions for inflation, such that there can be a nearly scale-invariant power spectrum $\cP_\cR(k)$ on large scales and a relatively small tensor-to-scalar ratio $r$, favored by the CMB observations \cite{Planck}.

Now, we calculate the power spectra ${\cal P}_{\cal R}(k)$, PBH abundances $f_\pb(M)$, and SIGW spectra $\Omega_{\rm GW}(f)$, respectively. The basic aims of the parameter adjustments in our model are threefold: to compose DM via PBHs, to generate sizable SIGW spectra, and to interpret the NANOGrav signal from the SIGW. Below, we explain these three aspects in more detail.

(1) For the PBHs in the two small-mass windows at $10^{-17}~M_\odot$ or $10^{-13}~M_\odot$, we demand the PBH abundance $f_\pb(M)$ to be 1, so as to compose all DM. However, if we wish to understand the NANOGrav signal via the SIGW that corresponds to the PBH of $30~M_\odot$, its abundance will be much smaller.

(2) We expect that the SIGW spectra $\Omega_{\rm GW}(f)$ are intense enough to reach the sensitivity curves of several next-generation GW detectors, such as Square Kilometer Array (SKA) \cite{ska}, International Pulsar Timing Array (IPTA) \cite{pta}, Laser Interferometer Space Antenna (LISA) \cite{lisa}, and Big Bang Observer (BBO) \cite{bbo}. However, at the same time, $\Omega_{\rm GW}(f)$ must avoid the constraints from the detectors on the run, such as advanced Laser Interferometer Gravitational-Wave Observatory (aLIGO) \cite{aligo}, since it has not observed GWs yet.

(3) We wish to interpret the potential isotropic stochastic GW background observed by the NANOGrav dataset via the SIGW. In Ref. \cite{NANO}, its latest 12.5-year analysis found strong evidence of a stochastic process, modeled as a power-law, with common amplitude and spectral slope across pulsars. For every process, it indicates the slope and amplitude at $1 \sg$ confidence level for the GW spectrum as
\begin{align}
\Omega_{\rm GWB}(f)=A_{\rm GWB}\lt(\f{f}{f_{\rm yr}}\rt)^\alpha, \label{app}
\end{align}
where $f_{\rm yr}=1~{\rm yr}^{-1}$ is the reference frequency, $A_{\rm GWB}$ is the amplitude at $f_{\rm yr}$, and $\alpha\in(-1.5,0.5)$ is the range of the slope of the potential isotropic stochastic GW background \cite{Yi:2021lxc} (illustrated as a blue parallelogram in Fig. \ref{fig:one30}). In this paper, we fix $\alpha$ to be its minimum $-1.5$ (i.e., the lowest slope) and demand the SIGW spectra $\Omega_{\rm GW}(f)$ to coincide with the top edge of the blue parallelogram. There are two basic reasons for this choice. First, the lowest slope provides the highest peak in $\Omega_{\rm GW}(f)$, making the SIGW most intense. Second, a higher $\Omega_{\rm GW}(f)$ corresponds to a higher PBH abundance, which will lead to stringent constraint on $f_{\rm PBH}(M)$, especially for the PBH of $30~M_\odot$.

\begin{figure*}[htb]
\begin{center}
\includegraphics[width=0.32\linewidth]{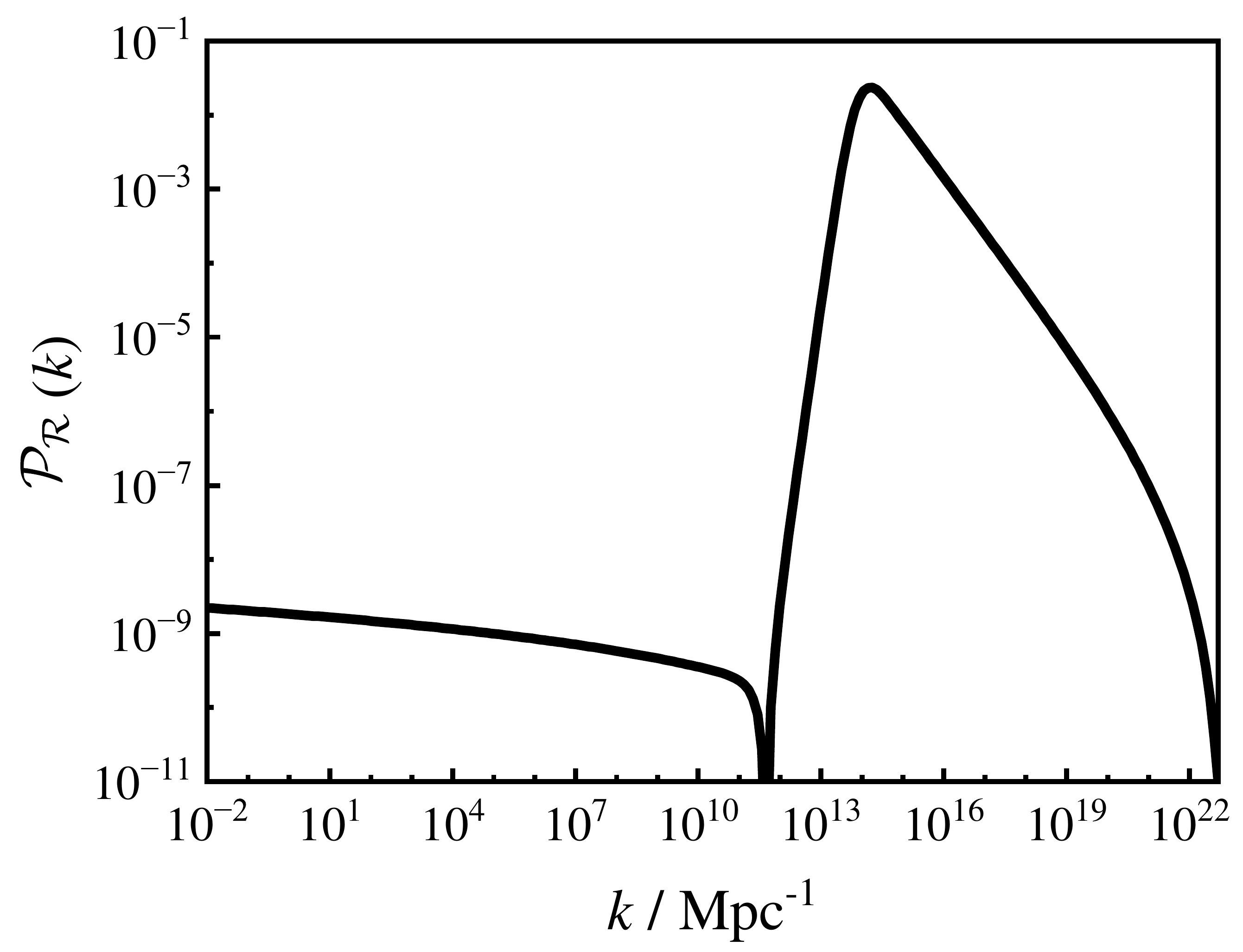}
\includegraphics[width=0.32\linewidth]{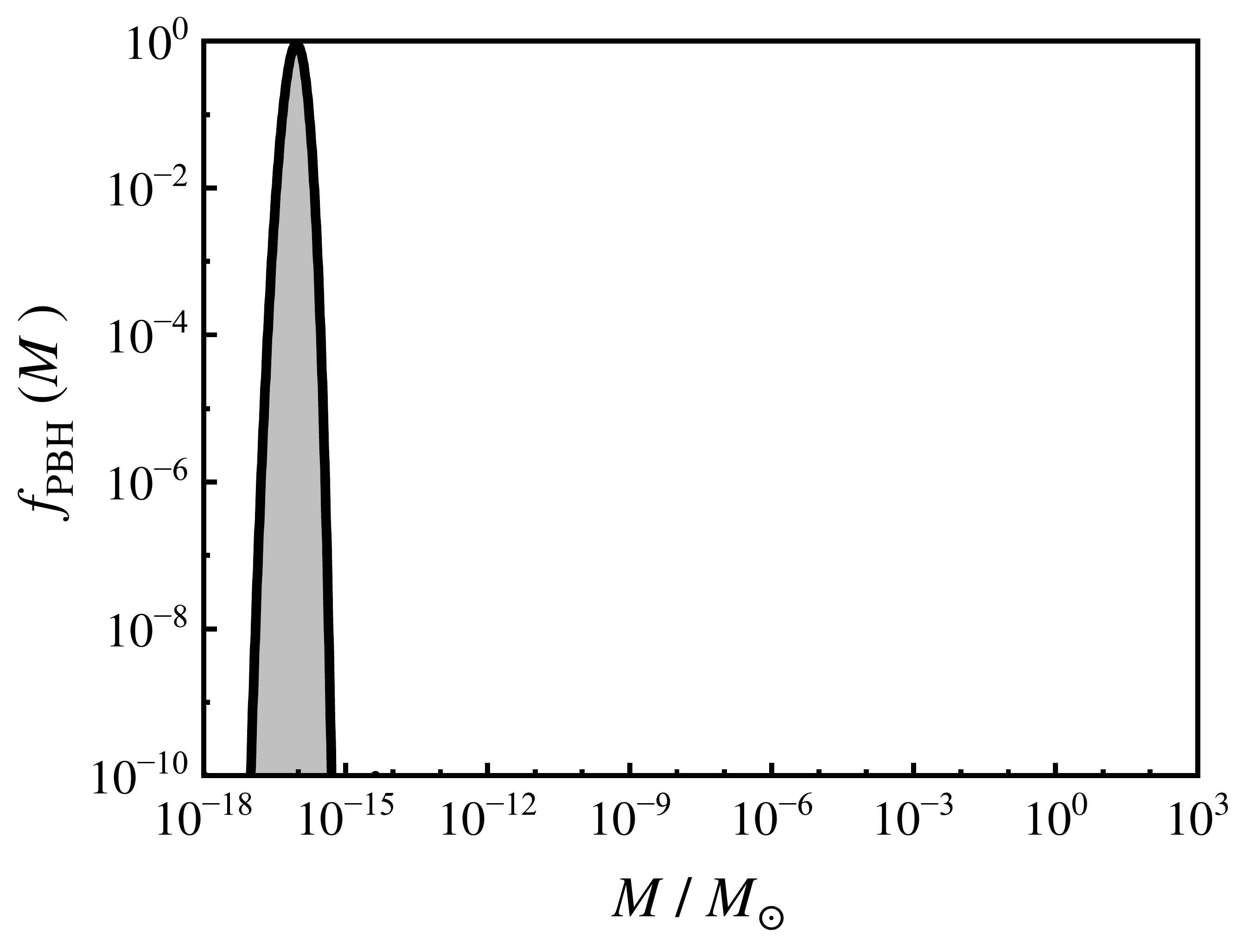}
\includegraphics[width=0.32\linewidth]{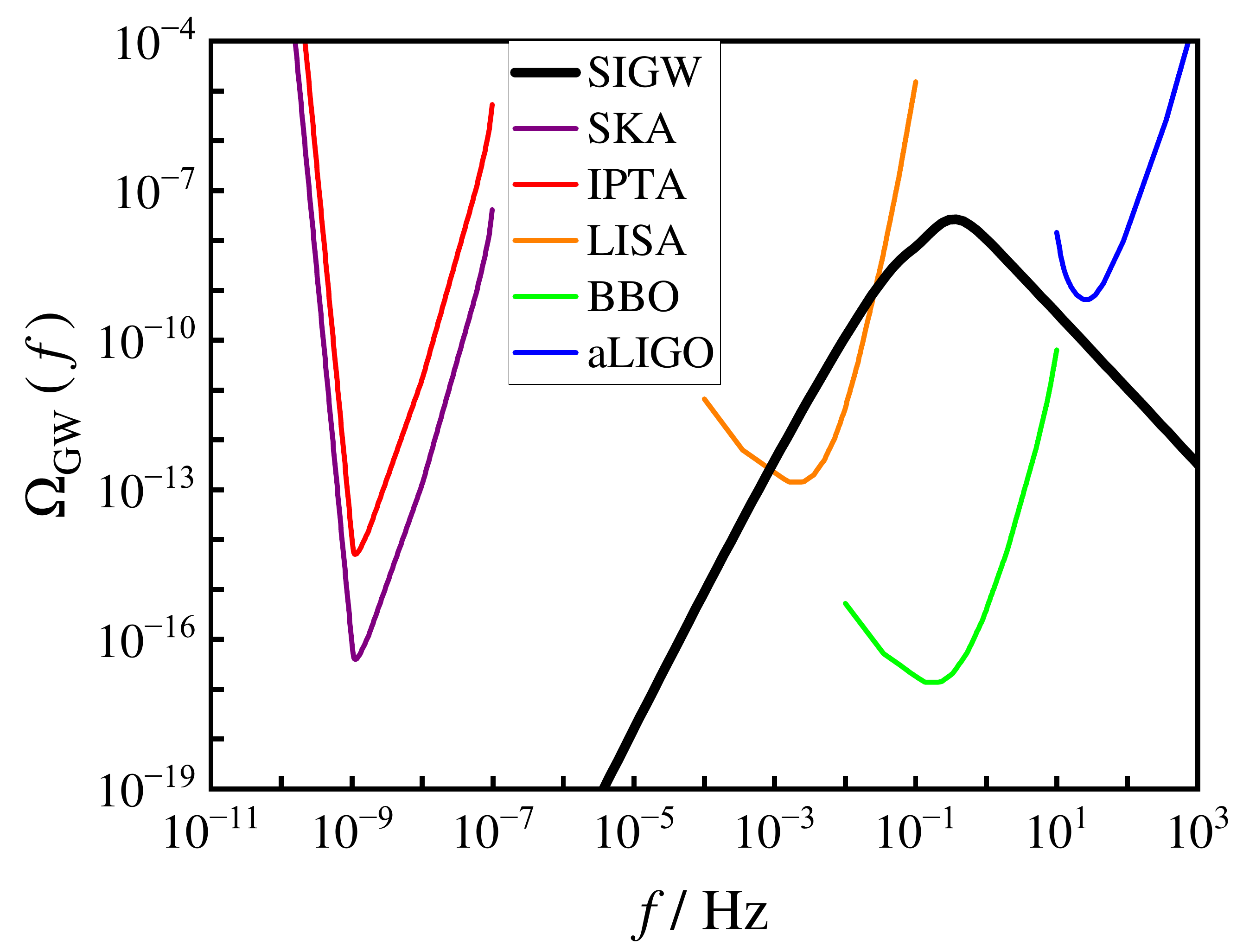}
\end{center}
\caption{The power spectrum ${\cal P}_{\cal R}(k)$, PBH abundance $f_{\rm PBH}(M)$, and SIGW spectrum $\Omega_{\rm GW}(f)$ with the PBH mass $M$ in the typical mass window at $10^{-17}~M_\odot$ (different sensitivity curves of various next-generation GW detectors are also presented in the right panel). The peak of ${\cal P}_{\cal R}(k)$ approaches around $10^{-2}$ in the USR inflation, remarkably enhancing $f_{\rm PBH}(M)$ and $\Omega_{\rm GW}(f)$ at certain mass and frequency. The PBH may have abundance $f_{\rm PBH}(M)\sim 1$ and thus compose all DM. The relevant SIGW is expected to reach the sensitivity curves of LISA and BBO, without touching the current constraint from aLIGO. Moreover, at low frequencies, the SIGW spectrum shows the $\Omega_{\rm GW}(f)\propto f^3$ scaling behavior.} \label{fig:one17}
\end{figure*}

\begin{figure*}[htb]
\begin{center}
\includegraphics[width=0.32\linewidth]{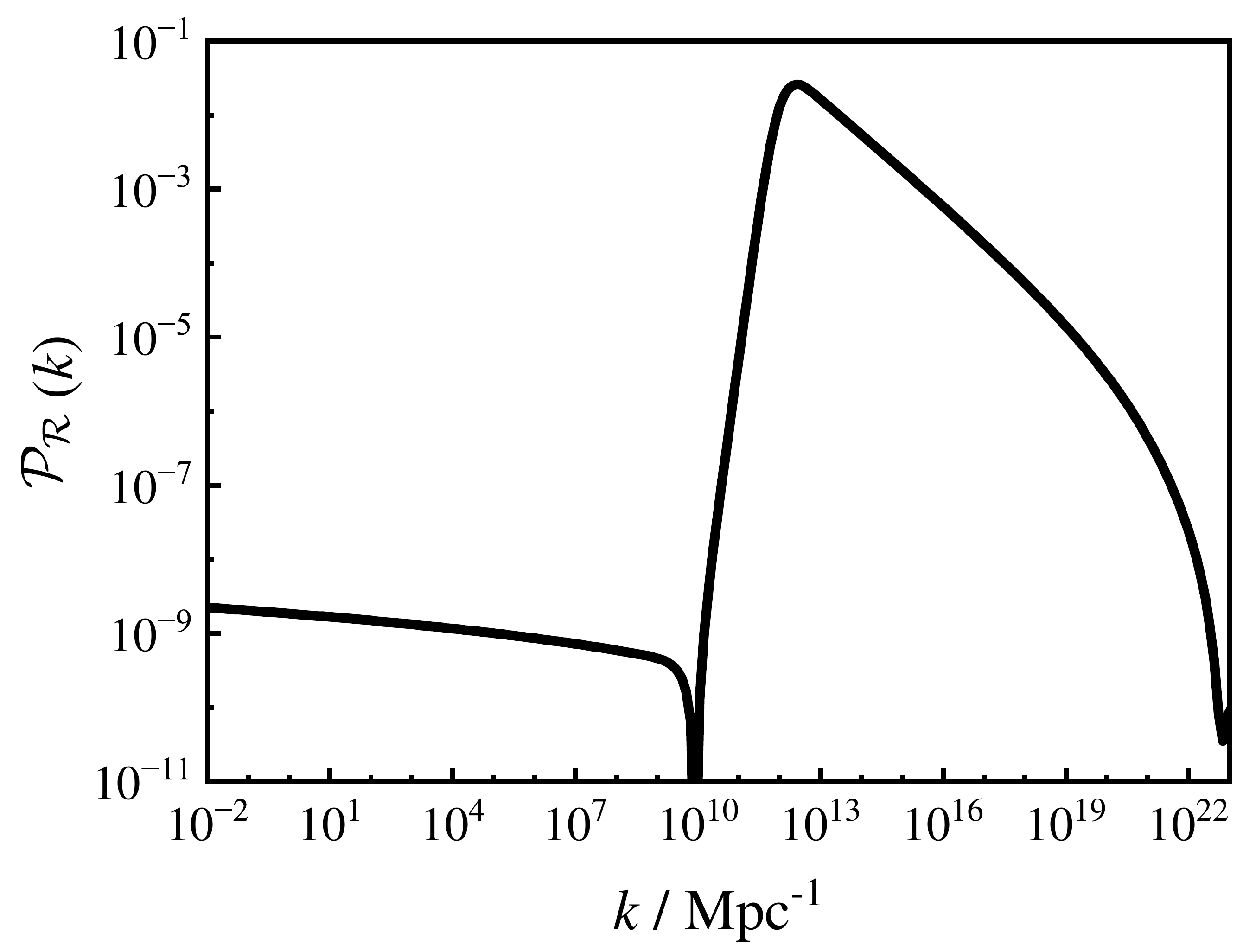}
\includegraphics[width=0.32\linewidth]{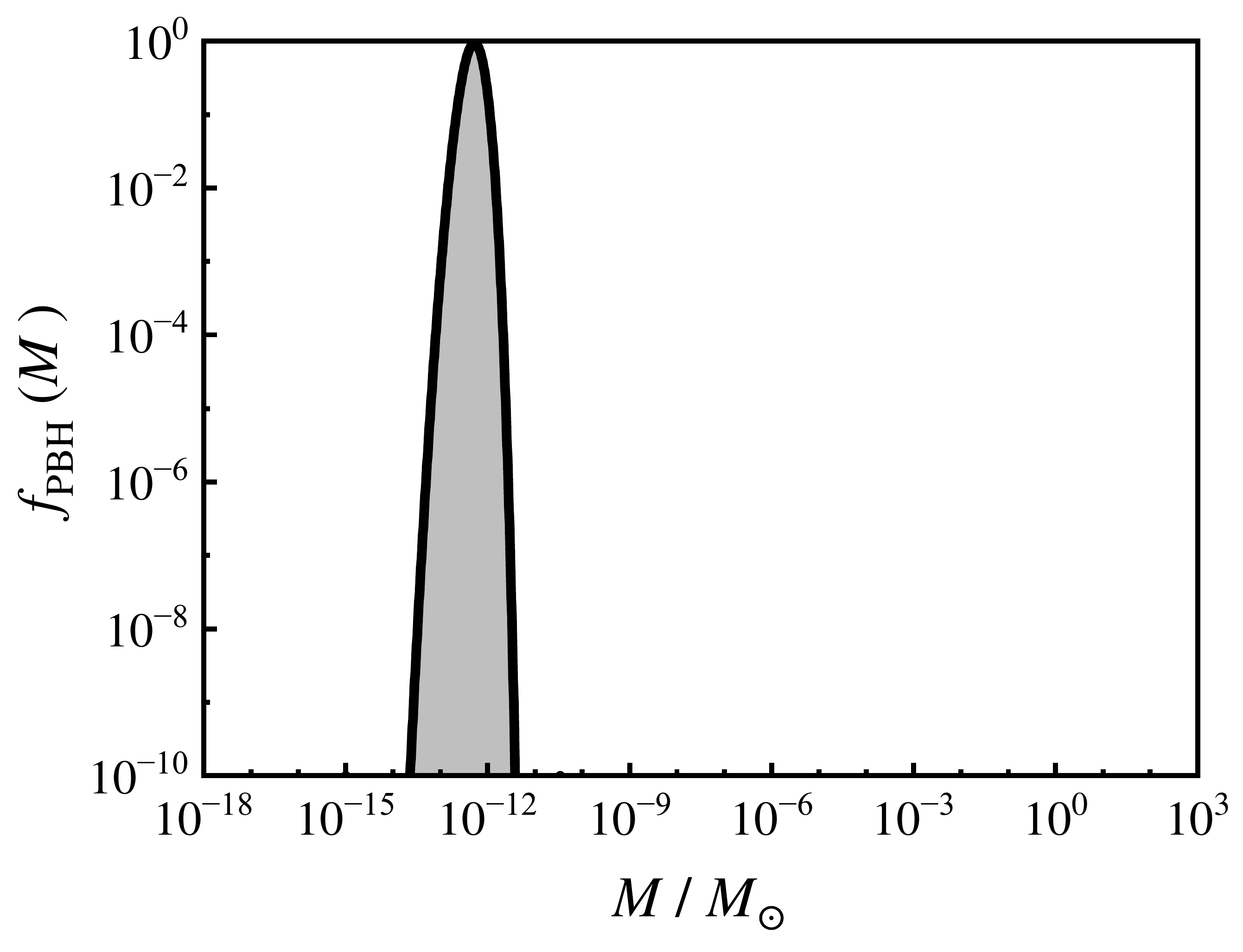}
\includegraphics[width=0.32\linewidth]{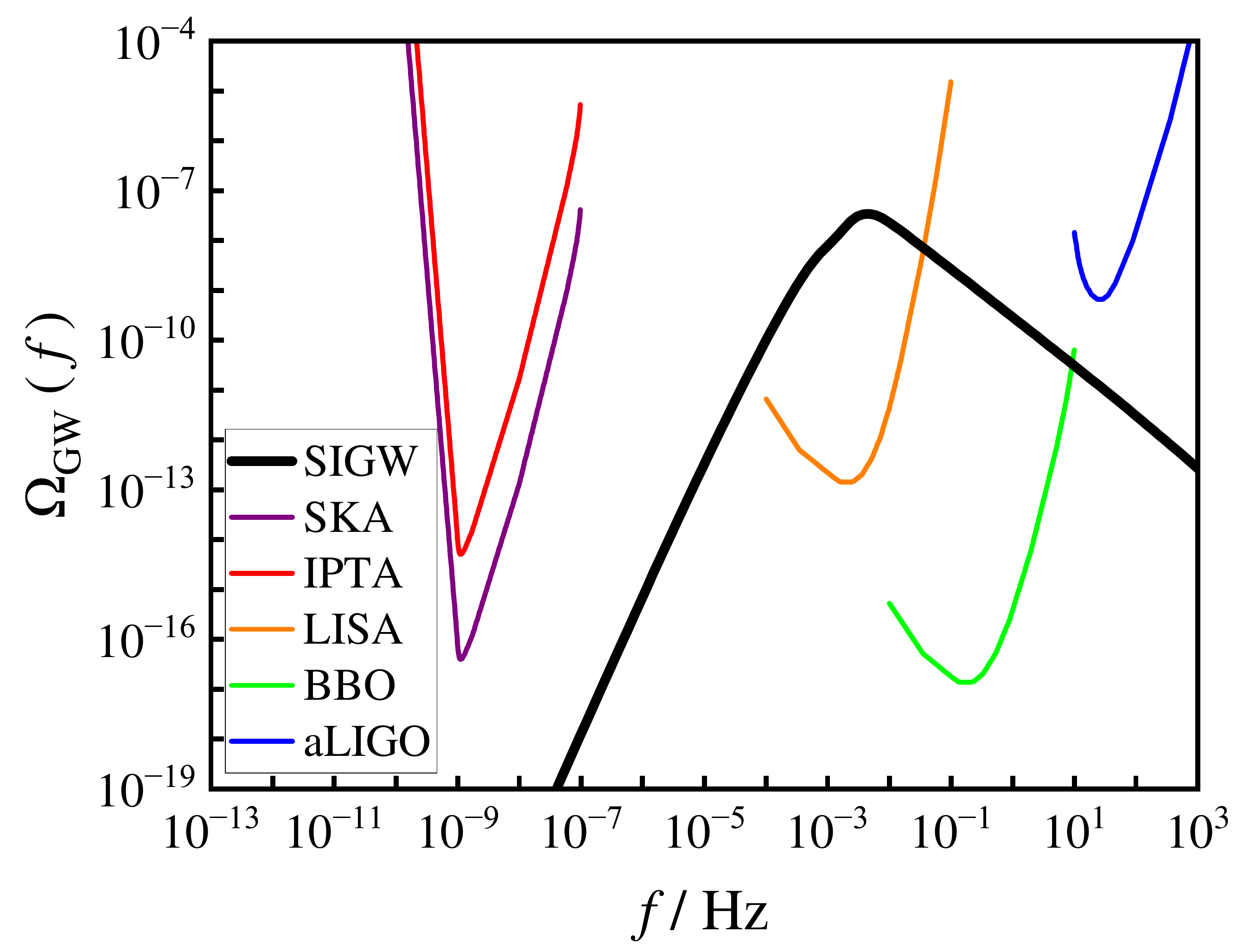}
\end{center}
\caption{Same as Fig. \ref{fig:one17}, but with the PBH mass $M$ in the typical mass window at $10^{-13}~M_\odot$. The PBH can still have $f_{\rm PBH}(M)\sim 1$ and thus compose all DM. Meanwhile, the peak of ${\cal P}_{\cal R}(k)$ moves to larger scale, and the peak of $\Omega_{\rm GW}(f)$ moves to lower frequency.} \label{fig:one13}
\end{figure*}

\begin{figure*}[htb]
\begin{center}
\includegraphics[width=0.32\linewidth]{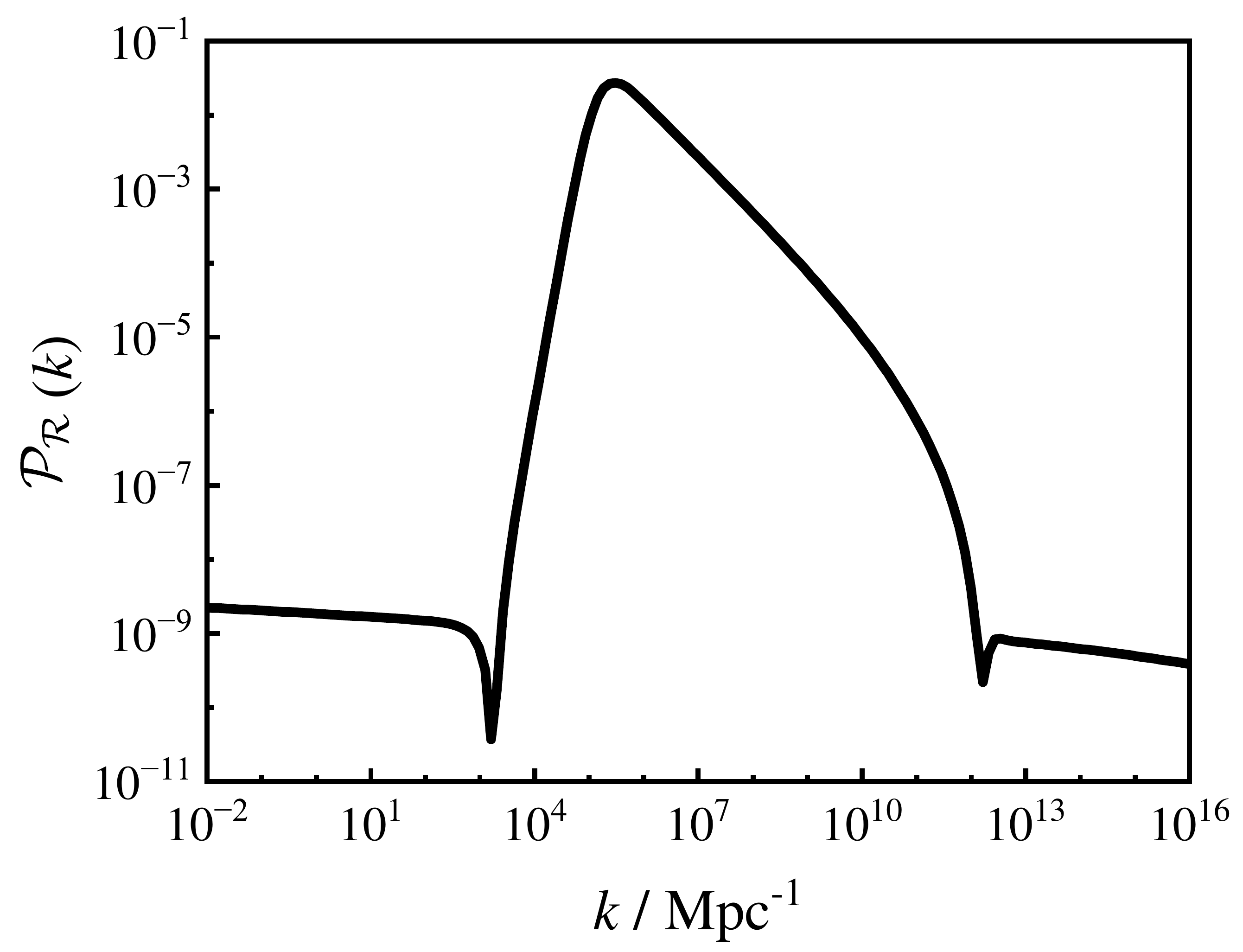}
\includegraphics[width=0.32\linewidth]{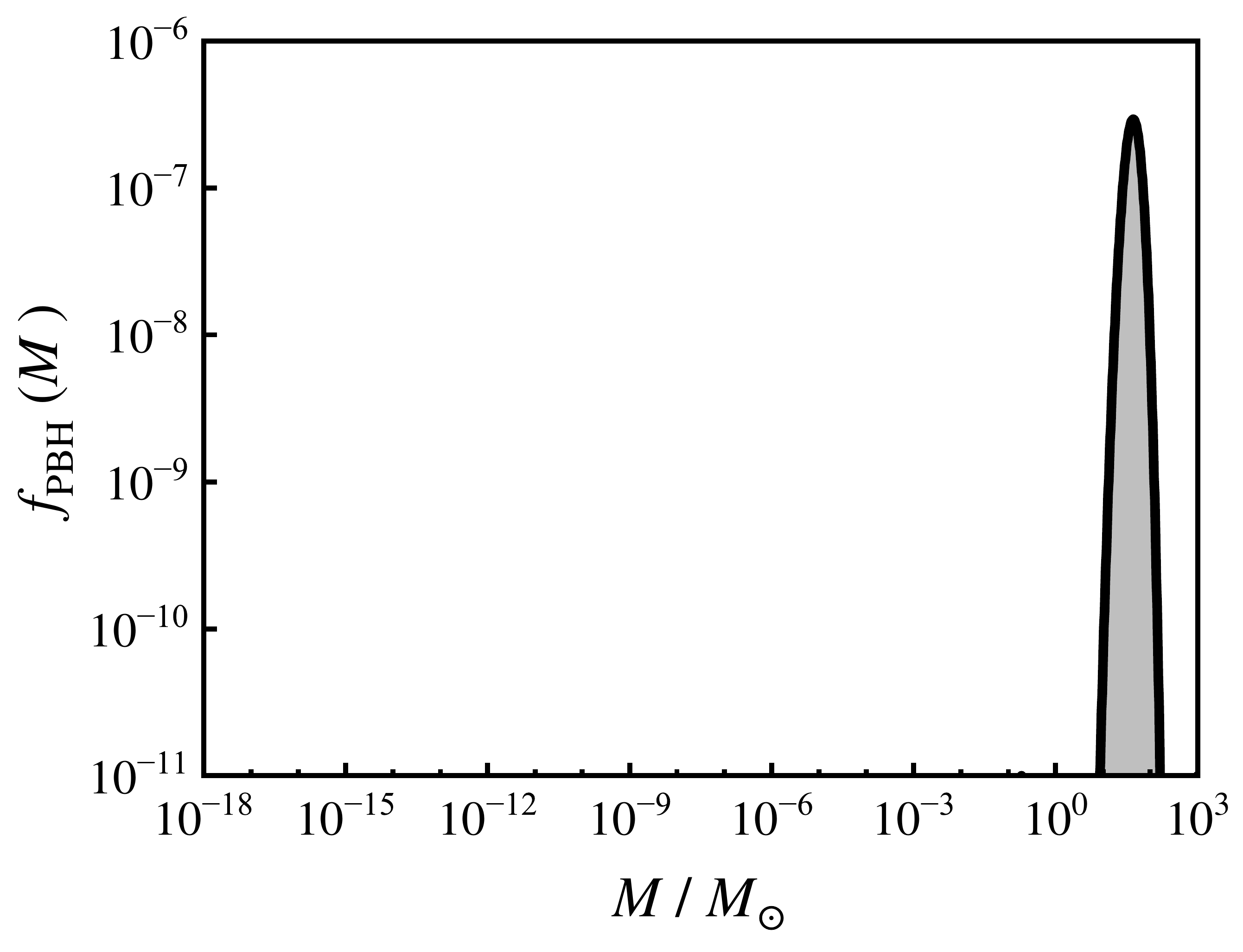}
\includegraphics[width=0.32\linewidth]{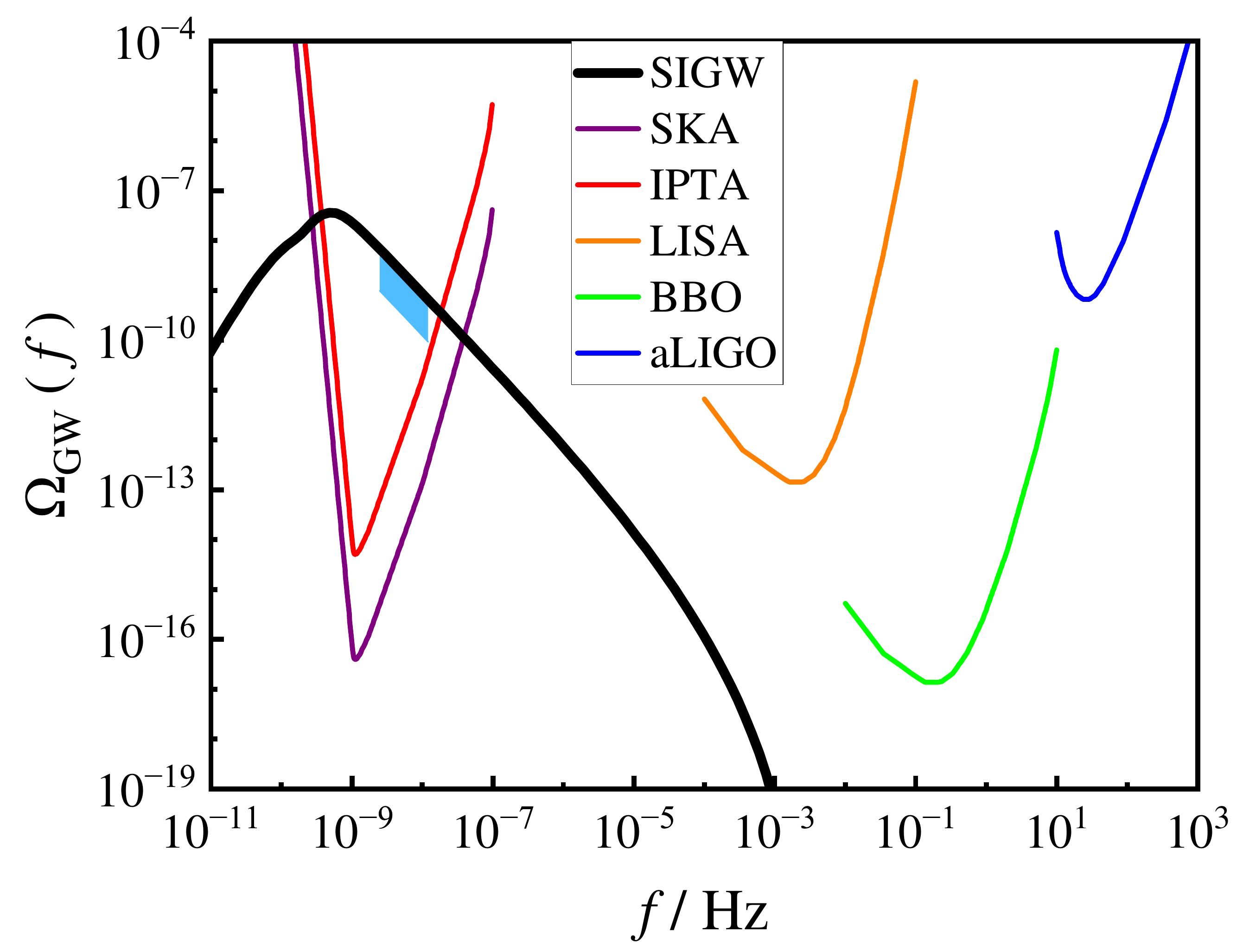}
\end{center}
\caption{Same as Fig. \ref{fig:one13}, but with the PBH mass $M$ in the typical mass window at $30~M_\odot$. The peak of ${\cal P}_{\cal R}(k)$ moves to even larger scale, and the peak of $\Omega_{\rm GW}(f)$ moves to even lower frequency. The SIGW spectrum is expected to reach the sensitivity curves of the next-generation GW detectors like SKA and IPTA. In this case, there are two notable characters, quite different from those in Figs. \ref{fig:one17} and \ref{fig:one13}. First, the SIGW can interpret the potential isotropic stochastic GW background from the NANOGrav 12.5-year dataset at $1\sg$ confidence level (blue parallelogram in the right panel). The slope $\alpha$ in Eq. (\ref{app}) is set to be $-1.5$ (i.e., the lowest slope), and the SIGW spectrum is tangent to the top edge of the blue parallelogram, so as to provide the highest peak of $\Omega_{\rm GW}(f)$. Second, the PBH abundance $f_\pb(M)$ is merely $10^{-7}$, meaning that even if the PBH of $30~M_\odot$ is definitely excluded as a candidate of DM, its relevant SIGW can still explain the NANOGrav signal.} \label{fig:one30}
\end{figure*}

Bearing the above three requirements in mind, we plot the power spectra ${\cal P}_{\cal R}(k)$, PBH abundances $f_\pb(M)$, and SIGW spectra $\Omega_{\rm GW}(f)$ in Figs. \ref{fig:one17}--\ref{fig:one30}. The model parameters $A_0$, $\phi_0$, and $\sg$ are summarized in Tab. \ref{tab:1}.

\begin{table}[htb]
\renewcommand\arraystretch{1.25}
\centering
\begin{tabular}{m{1.2cm}<{\centering}|m{1.8cm}<{\centering}|m{1.6cm}<{\centering}|m{1.6cm}<{\centering}}
\hline\hline
$M/M_\odot$ & $A_0$ & $\phi_0/m_{\rm P}$ & $\sigma/m_{\rm P}$\\
  \hline
$10^{-17}$  & 0.00264134 & 1.34  &  0.0944876 \\
  \hline
$10^{-13}$  & 0.003025   & 1.83  &  0.0451306 \\
\hline
$30$        & 0.00924    & 2.53  &  0.0172    \\
  \hline\hline
\end{tabular}
\caption{The parameters $A_0$, $\phi_0$, and $\sg$ for the PBH abundances $f_\pb(M)\sim 1$ in the two small-mass windows at $10^{-17}~M_\odot$ or $10^{-13}~M_\odot$, and $f_\pb(M)\sim 10^{-7}$ in the mass window at $30~M_\odot$. The SIGW spectra $\Omega_{\rm GW}(f)$ with these parameters can be observed by the next-generation GW detectors and avoid the current constraint. For the PBH of $30~M_\odot$, its relevant SIGW can interpret the potential isotropic stochastic GW background from the NANOGrav 12.5-year dataset.} \label{tab:1}
\end{table}

From Figs. \ref{fig:one17}--\ref{fig:one30} and Tab. \ref{tab:1}, our basic results can be drawn as follows.

(1) In the USR inflation, if the peak of the power spectrum ${\cal P}_{\cal R}(k)$ reaches $10^{-2}$ on small scales, the PBH abundance $f_\pb(M)$ is significantly enhanced, and the PBH can be considered as an effective candidate of DM. Simultaneously, the relevant SIGW spectrum $\Omega_{\rm GW}(f)$ is also enhanced at the corresponding frequency band and can be observed by different GW detectors in the future.

(2) With the peak of ${\cal P}_{\cal R}(k)$ moving to larger scales, the PBH mass $M$ increases, and the peak of $\Omega_{\rm GW}(f)$ moves to lower frequencies, as Eqs. (\ref{M}) and (\ref{f}) indicate that a smaller $k_\pb$ corresponds to a larger $M$ and a lower $f$. Also, a smaller $k_\pb$ means an earlier USR stage, so the parameter $\phi_0$ increases with $M$, as shown in Tab. \ref{tab:1}.

(3) As shown in Figs. \ref{fig:one17} and \ref{fig:one13}, the PBHs of $10^{-17}~M_\odot$ or $10^{-13}~M_\odot$ can compose all DM with $f_\pb(M)\sim 1$, and their relevant SIGW spectra $\Omega_{\rm GW}(f)$ are intense enough to reach the sensitivity curves of the next-generation GW detectors like LISA \cite{Bartolo:2018evs, Bartolo:2018rku, lisalisa1, lisalisa2} and BBO \cite{Kozaczuk:2021wcl, Gehrman:2022imk}, without touching the current constraint from aLIGO. Moreover, for the PBH of $30~M_\odot$, from the middle and right panels in Fig. \ref{fig:one30}, even if its abundance is strongly constrained to be merely $10^{-7}$ (the possibility as a stable candidate of DM is strictly excluded), its SIGW spectrum can still explain the potential isotropic stochastic GW background observed by the NANOGrav 12.5-year dataset \cite{DeLuca:2020agl, Vaskonen:2020lbd, Kohri:2020qqd, Inomata:2020xad, Yi:2022anu}. In addition, in Figs. \ref{fig:one17}--\ref{fig:one30}, the SIGW spectra show the universal infrared scaling behavior as $\Omega_{\rm GW}(f)\propto f^3$, in agreement with Refs. \cite{Yuan:2019wwo, Liu:2020oqe, Cai:2019cdl, Hook:2020phx}.

Last, we discuss some details in the adjustments of the three parameters $A_0$, $\phi_0$, and $\sg$ in our model. As shown in Refs. \cite{liuyichen, wangqing}, if we only focus on the PBH abundance $f_\pb(M)$, two parameters $\phi_0$ and $\sg$ are already sufficient. However, in the present work, we also pay attention to the SIGW spectrum $\Omega_{\rm GW}(f)$, so the third parameter $A_0$ is indispensable. Amongst them, the width $\sg$ is the most influential factor, as it strongly affects the profile of $\Omega_{\rm GW}(f)$. For instance, for the PBHs of $10^{-17}~M_\odot$ or $10^{-13}~M_\odot$, we demand a steeper $\Omega_{\rm GW}(f)$, so that it does not contradict the current constraint from aLIGO, and for the PBH of $30~M_\odot$, we need $\Omega_{\rm GW}(f)$ to possess the lowest slope $-1.5$ in its decreasing region to interpret the NANOGrav signal in the right panel in Fig. \ref{fig:one30}. Moreover, $A_0$ plays a similar role as $\sigma$ in calculating $f_\pb(M)$, so it helps to break the parameter degeneracy. Nevertheless, due to the antisymmetric form of the perturbation $\dt V$, the fine-tuning problem frequently met in the USR inflation has already been relieved greatly.

\section{PBHs and SIGWs from two perturbations on the inflaton potential} \label{sec:two}

In this section, we further investigate the cases with two perturbations on the background inflaton potential, so that there can be PBHs of different masses with appropriate abundances in two of the three typical mass windows at $10^{-17}~M_\odot$, $10^{-13}~M_\odot$, and $30~M_\odot$ simultaneously. Moreover, the relevant SIGW spectra are also explored in detail.

\begin{figure*}[htb]
\centering
\includegraphics[width=0.32\linewidth]{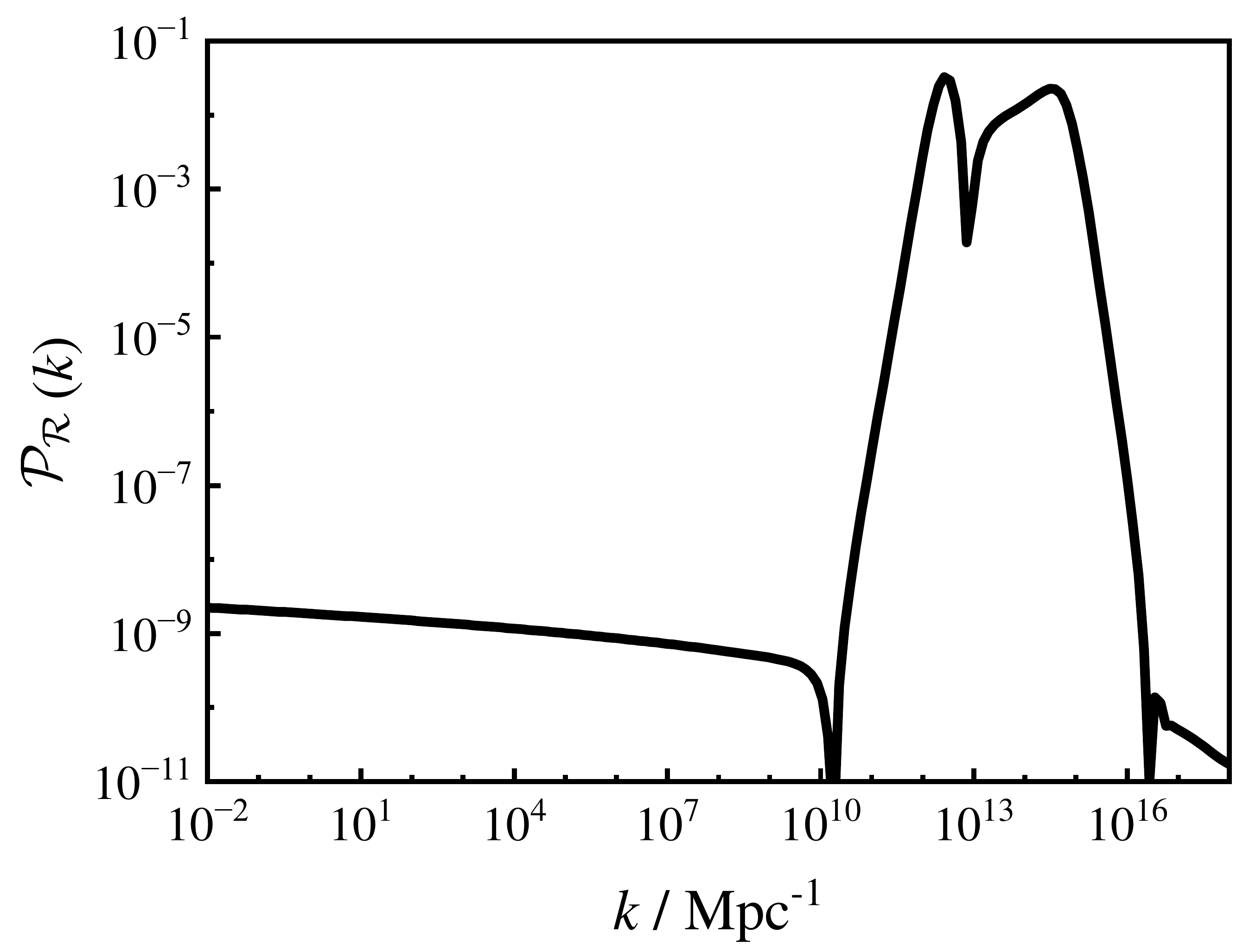}
\includegraphics[width=0.32\linewidth]{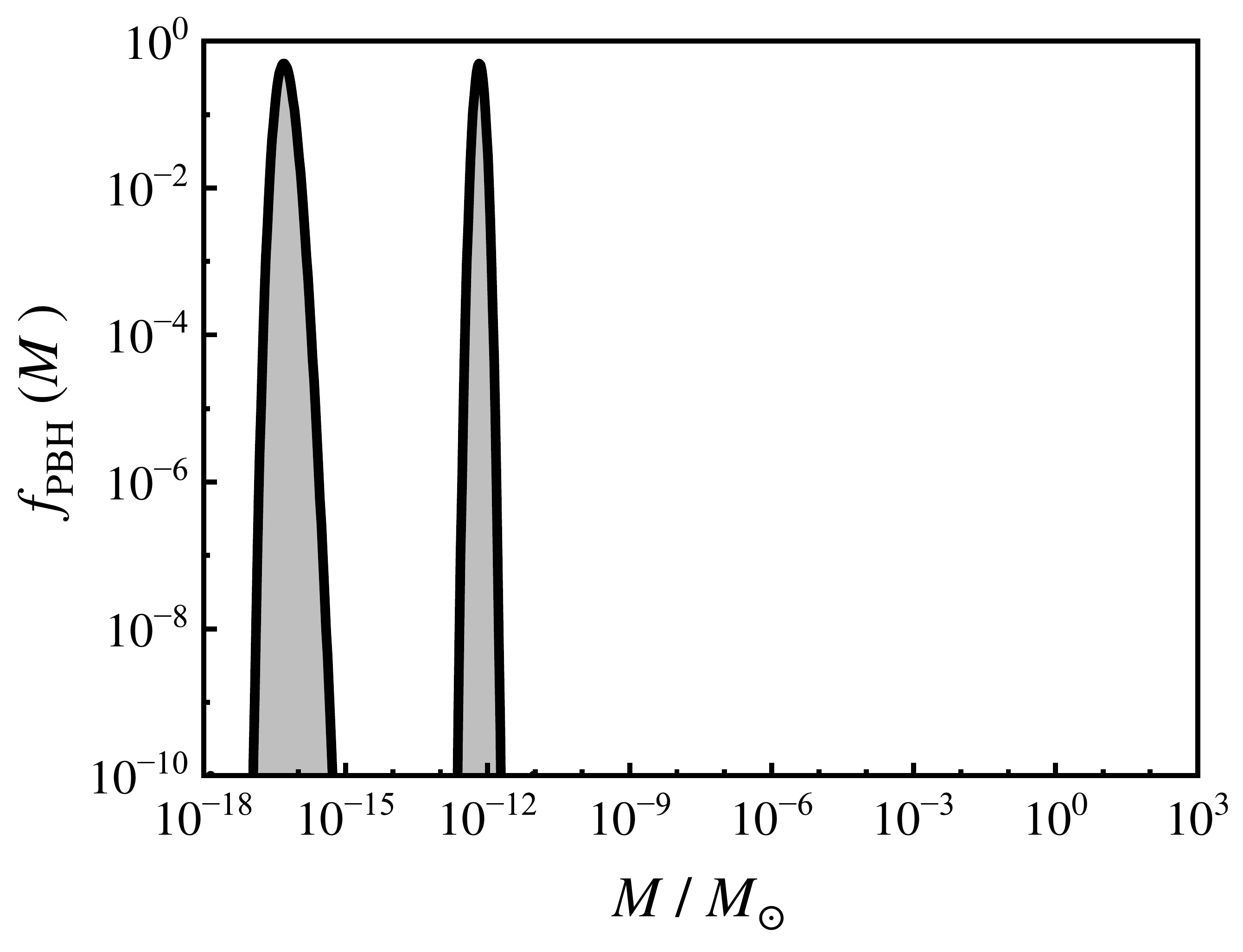}
\includegraphics[width=0.32\linewidth]{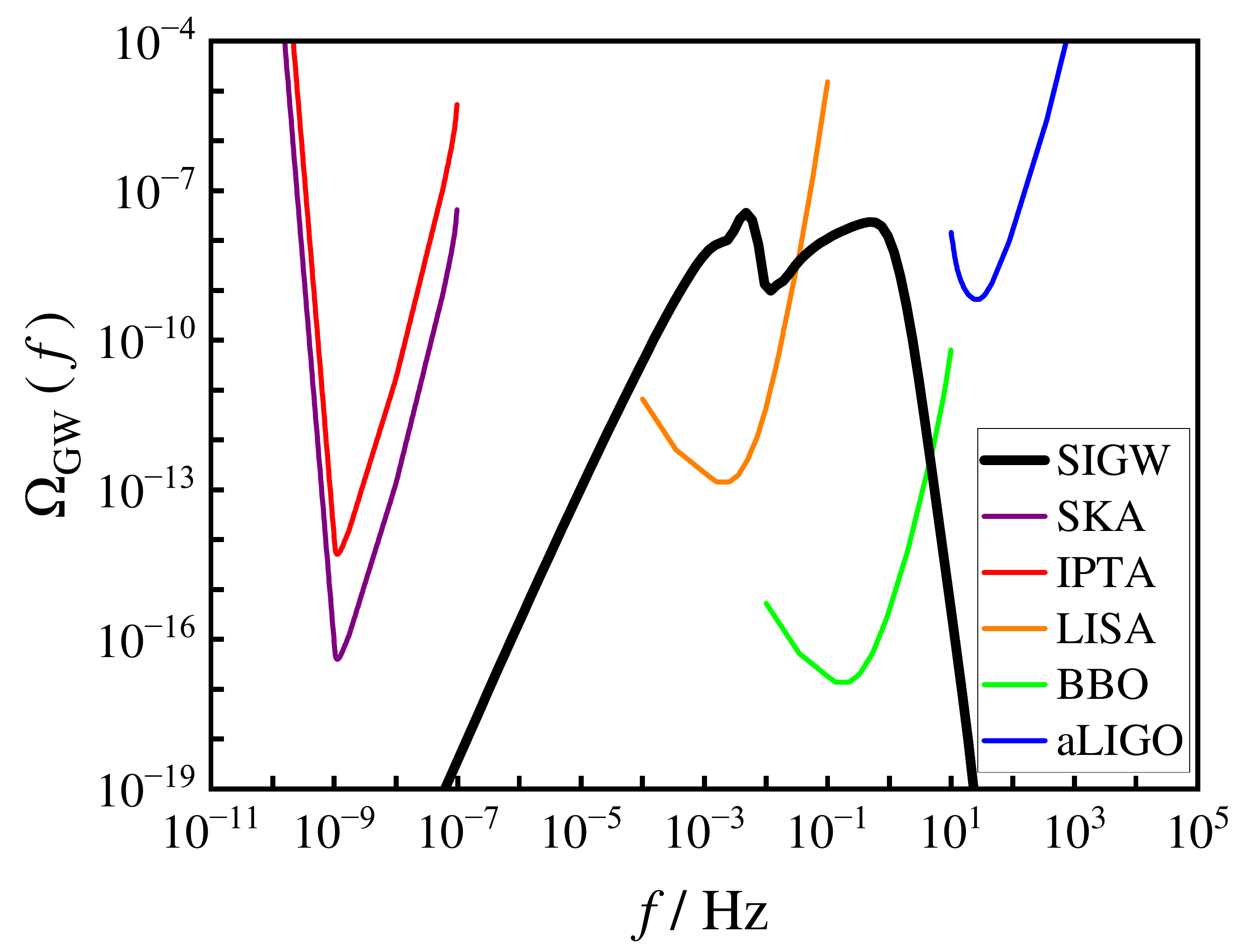}
\caption{The power spectrum ${\cal P}_{\cal R}(k)$, PBH abundance $f_\pb(M)$, and SIGW spectrum $\Omega_{\rm GW}(f)$ with the PBH masses in two typical mass windows at $10^{-17}~M_\odot$ and $10^{-13}~M_\odot$. Both PBH abundances are set to be $0.5$, so that the PBHs can compose all DM. The relevant SIGW is expected to be observed by LISA and BBO, without touching the current constraint from aLIGO. Some distortions appear in ${\cal P}_{\cal R}(k)$ and $\Omega_{\rm GW}(f)$, because there is inevitable interference between the two USR stages.} \label{fig:two1713}
\end{figure*}

\begin{figure*}[htb]
\centering
\includegraphics[width=0.32\linewidth]{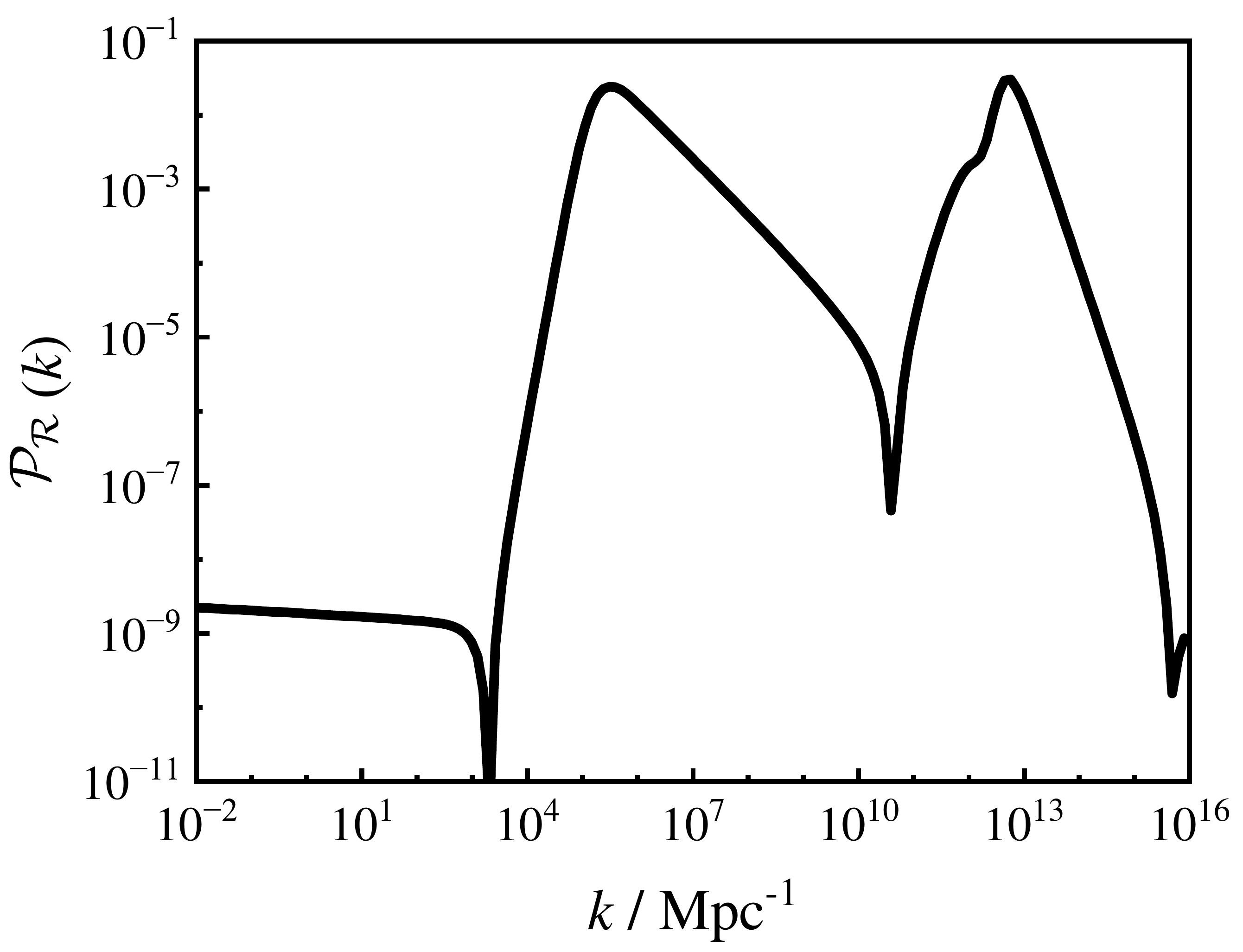}
\includegraphics[width=0.32\linewidth]{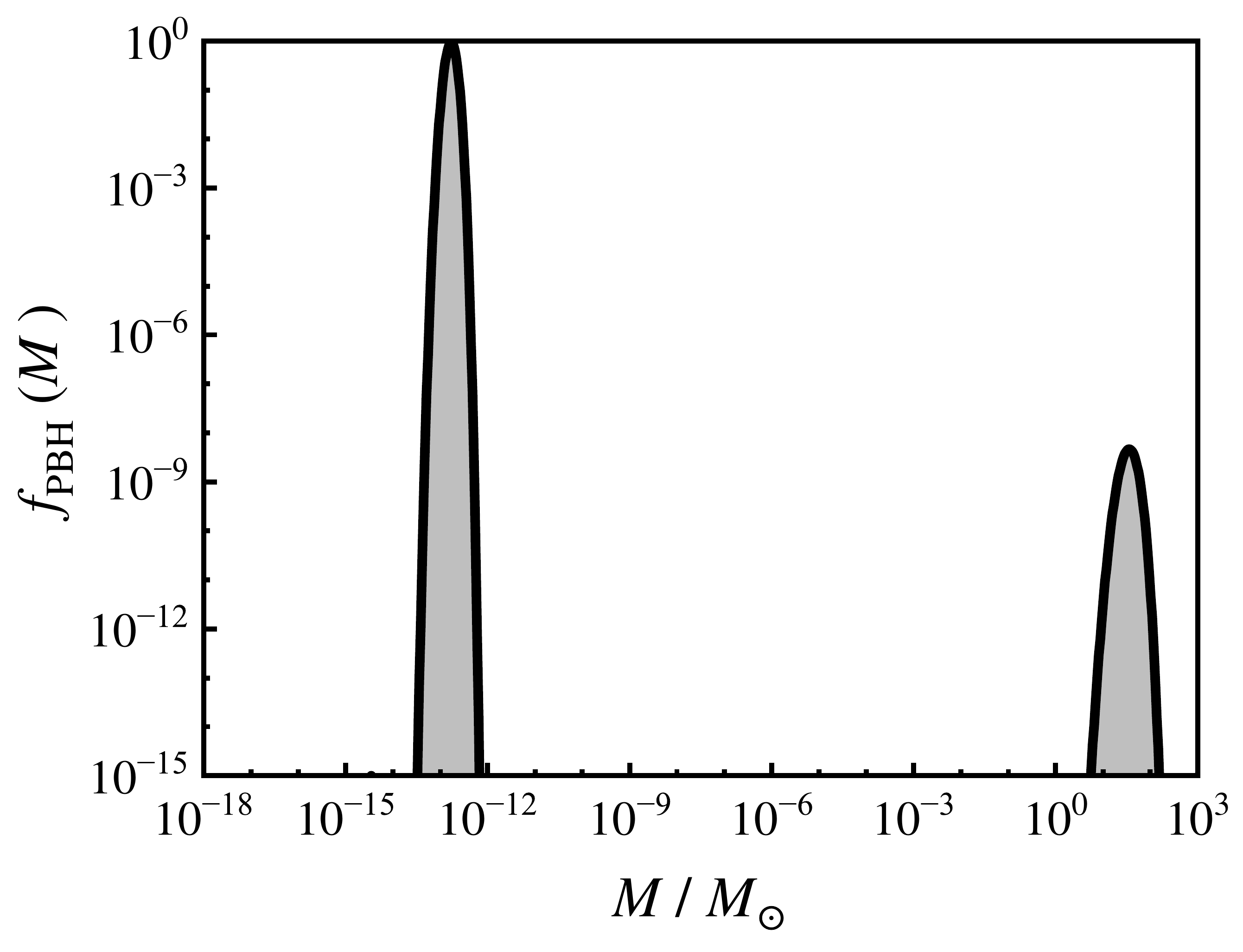}
\includegraphics[width=0.32\linewidth]{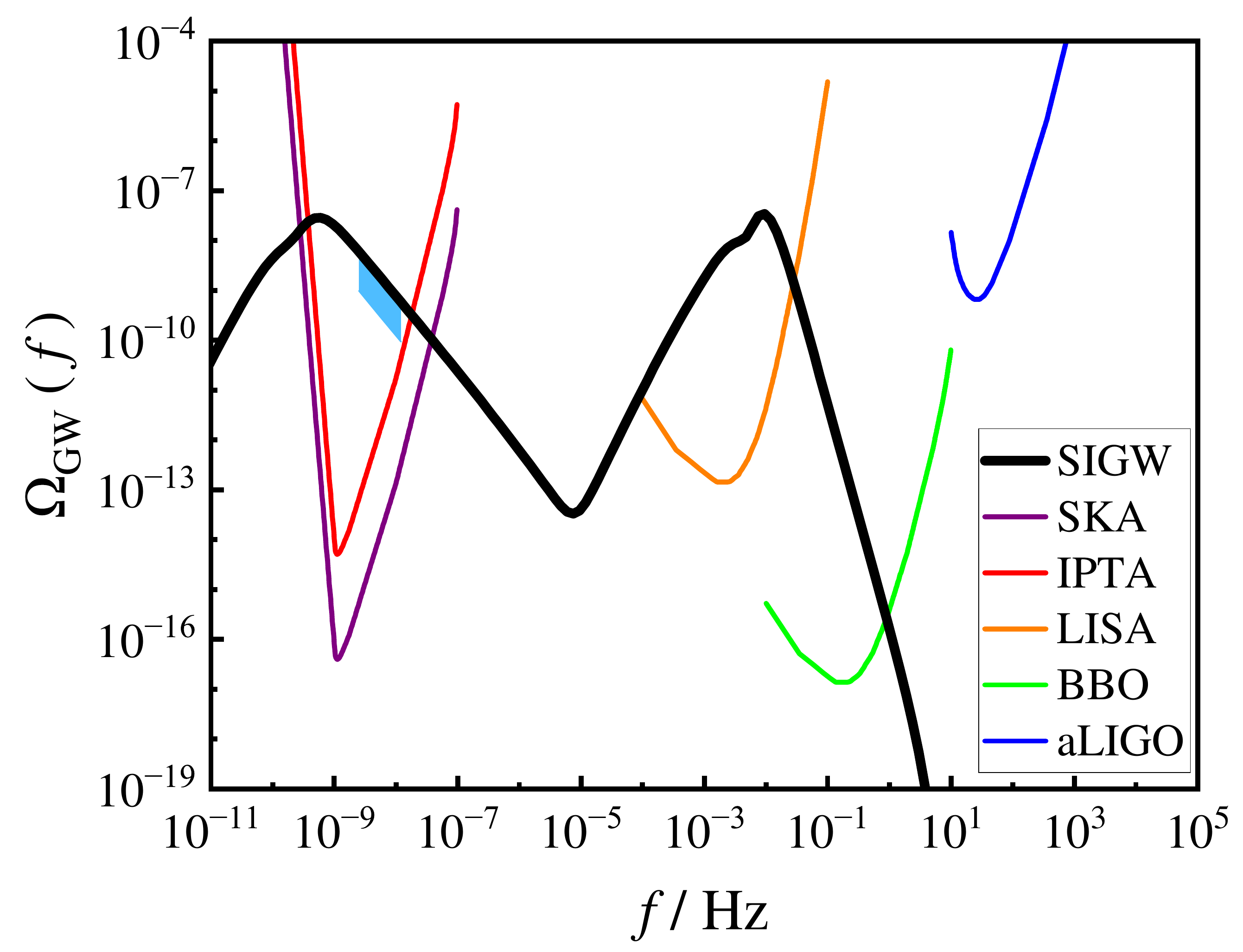}
\caption{Same as Fig. \ref{fig:two1713}, but with the PBH masses in two typical mass windows at $10^{-13}~M_\odot$ and $30~M_\odot$. The SIGW is expected to be observed by IPTA, SKA, LISA, and BBO, but not by aLIGO. Moreover, the NANOGrav signal is shown as a blue parallelogram in the right panel (as explained in Fig. \ref{fig:one30}), and the SIGW spectrum is tangent to its top edge. This condition strictly constrains the abundance of the PBH of $30~M_\odot$ to be merely $10^{-9}$ and thus excludes its possibility as a candidate of DM. Meanwhile, the abundance of the PBH of $10^{-13}~M_\odot$ is set to be 1, so that it can compose all DM alone.} \label{fig:two1330}
\end{figure*}

\begin{figure*}[htb]
\centering
\includegraphics[width=0.32\linewidth]{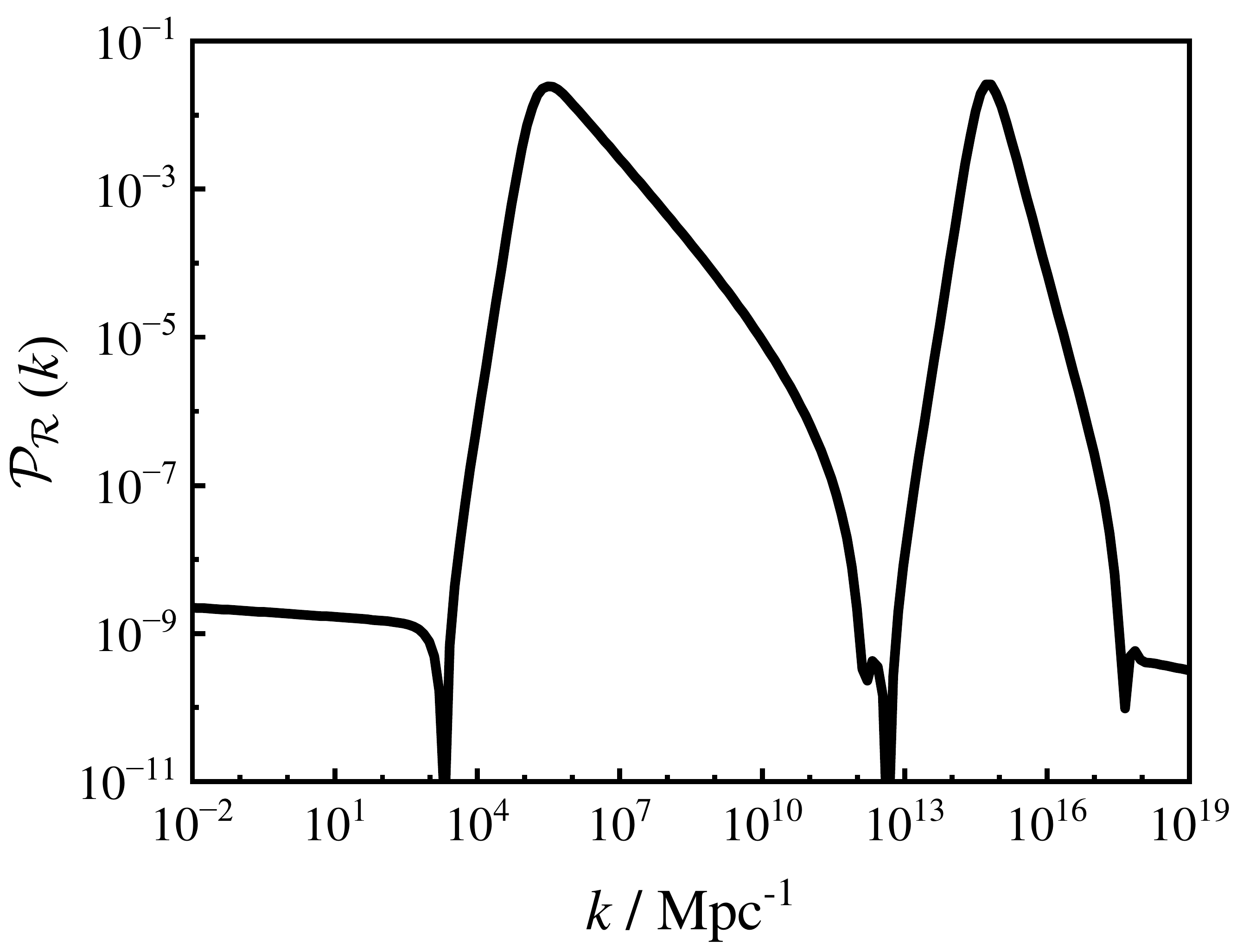}
\includegraphics[width=0.32\linewidth]{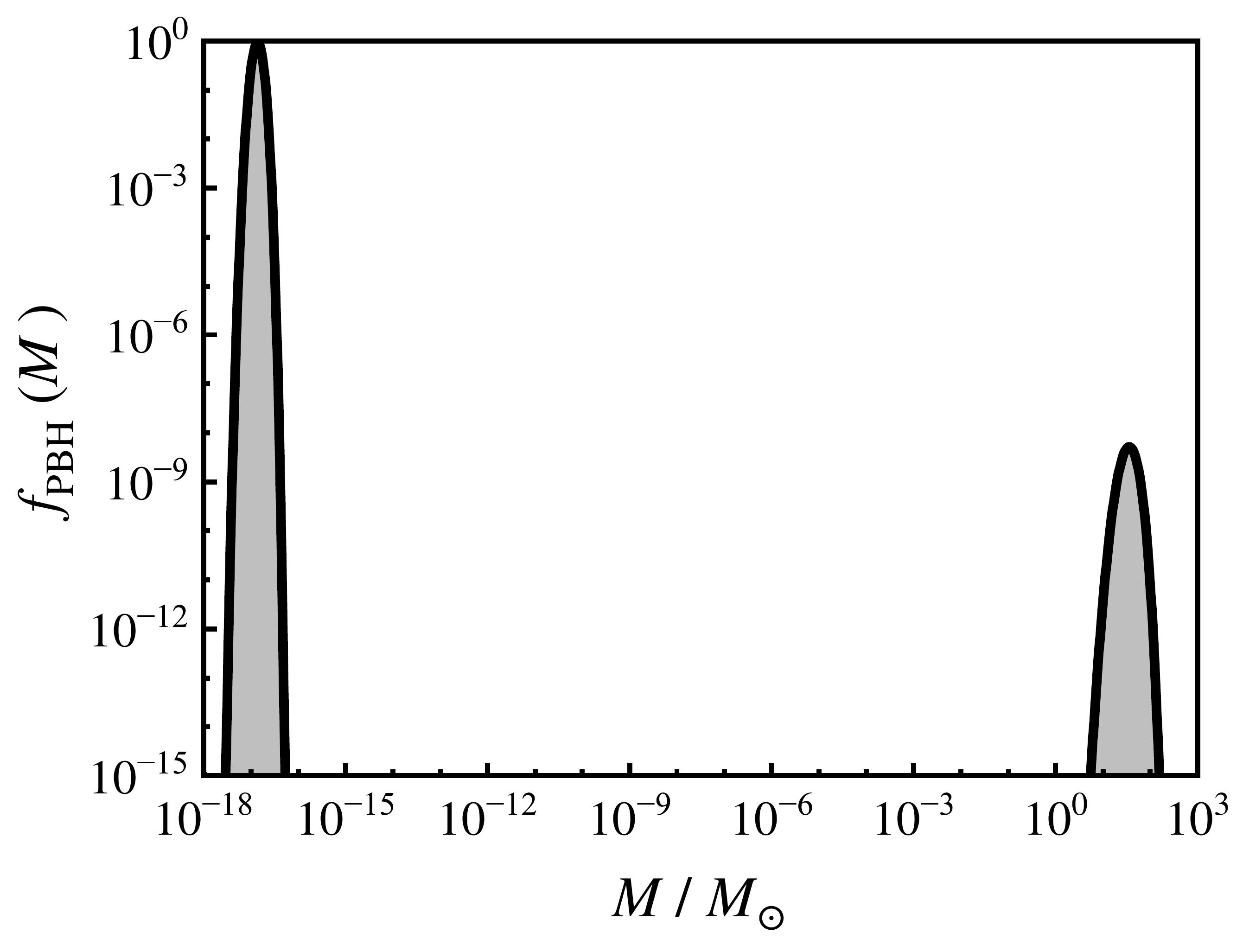}
\includegraphics[width=0.32\linewidth]{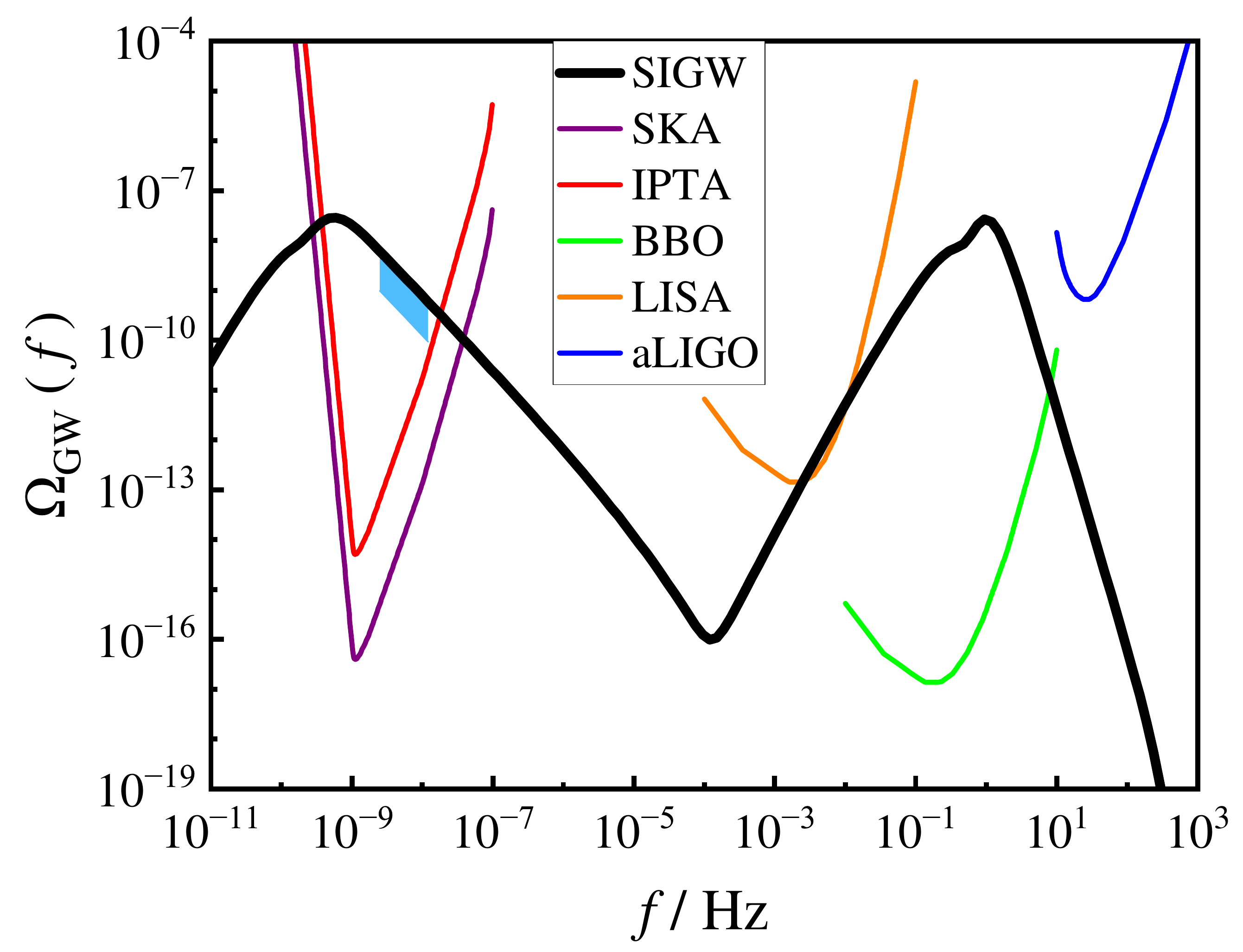}
\caption{Same as Fig. \ref{fig:two1330}, but with the PBH masses in two typical mass windows at $10^{-17}~M_\odot$ and $30~M_\odot$. The abundance of the PBH of $10^{-17}~M_\odot$ is set to be 1 to compose all DM, and the abundance of the PBH of $30~M_\odot$ is only $10^{-9}$, but is already sufficient to explain the NANOGrav signal.} \label{fig:two1730}
\end{figure*}

Now, the inflaton potential reads $V(\phi)=V_{\rm b}(\phi)+\dt V_1(\phi)+\dt V_2(\phi)$, and the two perturbations possess the same form as that in Eq. (\ref{3para}). Hence, there are six model parameters at present: $A_0^{(1)}$, $\phi_0^{(1)}$, $\sigma^{(1)}$, $A_0^{(2)}$, $\phi_0^{(2)}$, and $\sigma^{(2)}$ (the superscripts 1 and 2 stand for small and large PBH masses). According to the separation between the two PBH masses, the power spectra ${\cal P}_{\cal R}(k)$, PBH abundances $f_\pb(M)$, and SIGW spectra $\Omega_{\rm GW}(f)$ are plotted in Figs. \ref{fig:two1713}--\ref{fig:two1730}, and the corresponding model parameters are summarized in Tab. \ref{tab:2}. The initial conditions for inflation are kept the same as those in Sec. \ref{sec:one}.

\begin{table*}[htb]
\renewcommand\arraystretch{1.25}
\centering
\resizebox{1\columnwidth}{!}{
\begin{tabular}{m{2.8cm}<{\centering}| m{1.5cm}<{\centering}| m{1.3cm}<{\centering}| m{1.5cm}<{\centering}| m{1.5cm}<{\centering}| m{1.3cm}<{\centering}| m{1.5cm}<{\centering}| m{2.7cm}<{\centering}}
\hline\hline
$M/M_\odot$ & $A_0^{(1)}$ & $\phi_0^{(1)}/\m$ & $\sigma^{(1)}/\m$ & $A_0^{(2)}$ & $\phi_0^{(2)}/\m$ & $\sigma^{(2)}/\m$ & $(\phi_0^{(2)}-\phi_0^{(1)})/\m$ \\
  \hline
$10^{-17}$ and $10^{-13}$ & 0.3477753 & 1.720 & 0.029705  & 0.2726004 & 1.810 & 0.0290004 & 0.090 \\
  \hline
$10^{-13}$ and $30$         & 0.0788911 & 2.450 & 0.0146459 & 0.0093681 & 2.527 & 0.0172428 & 0.077 \\
  \hline
$10^{-17}$ and 30       & 0.0797032 & 2.280 & 0.0166578 & 0.0093281 & 2.527 & 0.0172428 & 0.247 \\
\hline\hline
\end{tabular}}
\caption{The parameters $A_0^{(1)}$, $\phi_0^{(1)}$, $\sigma^{(1)}$, $A_0^{(2)}$, $\phi_0^{(2)}$, and $\sigma^{(2)}$ (superscripts 1 and 2 stand for small and large PBH masses) for the PBHs with the appropriate abundances in two of the three typical mass windows at $10^{-17}~M_\odot$, $10^{-13}~M_\odot$, and $30~M_\odot$. The SIGW spectra with these parameters are expected to be observed by the next-generation GW detectors and to avoid the current constraint. For the cases involving the PBH of $30~M_\odot$, the relevant SIGWs can also explain the potential isotropic stochastic GW background from the NANOGrav 12.5-year dataset.} \label{tab:2}
\end{table*}

From Figs. \ref{fig:two1713}--\ref{fig:two1730} and Tab. \ref{tab:2}, we arrive at the following results.

(1) Because of the two perturbations on the background inflaton potential, there appear two peaks in ${\cal P}_{\cal R}(k)$, $f_\pb(M)$, and $\Omega_{\rm GW}(f)$ simultaneously. Analogous to the cases with one perturbation, the PBHs can compose all DM, and the SIGWs are expected to be observed by the next-generation GW detectors or to interpret the NANOGrav signal, as long as the power spectra $\cP_\cR(k)$ are enhanced up to $10^{-2}$ on the relevant scales.

(2) In Fig. \ref{fig:two1713}, for the PBHs of $10^{-17}~M_\odot$ and $10^{-13}$ $M_\odot$, their abundances are both set to be $0.5$, in order to compose all DM. However, in Fig. \ref{fig:two1330}, for the PBHs of $10^{-13}~M_\odot$ and $30~M_\odot$, only the abundance of the former is set to be 1, as the abundance of the latter is negligibly small (around $10^{-9}$), but this is already enough to explain the potential isotropic stochastic GW background from the NANOGrav 12.5-year dataset. The situation is similar for the case with the PBHs of $10^{-17}~M_\odot$ and $30~M_\odot$.

(3) In Tab. \ref{tab:2}, one may naively expect that the separation between the two perturbations $\phi_0^{(2)}-\phi_0^{(1)}$ increases with the mass difference between the two mass windows, but this is not always the case. There are two basic reasons for this point. First, the two perturbations cannot be too close. Otherwise, there will be strong parameter degeneracy. Second, they cannot be too far away, either. Otherwise, the inflaton will spend much time on the first plateau and will pass the second one at much later times, making the relevant PBH mass extremely small. Therefore, the two perturbations should be placed at a moderate distance, and this will lead to inevitable interference between them accordingly. As a result, some distortions appear in ${\cal P}_{\cal R}(k)$ and $\Omega_{\rm GW}(f)$, as shown in Figs. \ref{fig:two1713}--\ref{fig:two1730}. This is a natural consequence from the overlap between the decaying and growing regions of ${\cal P}_{\cal R}(k)$ on different scales or $\Omega_{\rm GW}(f)$ on different frequencies.
Hence, merely adjusting the parameters $\sigma^{(1)}$ and $\sigma^{(2)}$ is not enough, and the parameters $A_0^{(1)}$ and $A_0^{(2)}$ are indispensable to alleviate the parameter degeneracy.

\section{Conclusion} \label{sec:con}

Since the discovery of the merging GWs from binary black holes in recent years, PBHs and SIGWs have aroused continuously growing research enthusiasm. One of the basic motivations is to compose DM via PBHs, and their relevant SIGWs are expected to be observed by the next-generation GW detectors. Hence, the aim of this paper is to study the PBH abundance and SIGW spectrum phenomenologically, by introducing the perturbations $\dt V$ on the background inflaton potential $V_{\rm b}(\phi)$. We systematically calculate the power spectra $\cP_\cR(k)$, PBH abundances $f_\pb(M)$, and SIGW spectra $\Omega_{\rm GW}(f)$ via the GLMS approximation of peak theory. We also wish to explain the potential isotropic stochastic GW background detected by the NANOGrav 12.5-year dataset. Our basic conclusions are summarized as follows.

(1) We choose the antisymmetric perturbation with three model parameters $A_0$, $\phi_0$, and $\sg$, so as to construct a plateau flat enough on $V_{\rm b}(\phi)$, leading inflation into the USR stage. The perturbation of this form can be connected to $V_{\rm b}$ very smoothly, not only relieving the fine-tuning problems that usually appear in the parameter adjustments, but also making $f_\pb(M)$ spike-like and avoiding the over-production of the PBHs of tiny masses, which circumvents the constraint from the extra-galactic gamma-ray bursts \cite{zs}.

(2) The USR inflation can dramatically enhance $\cP_\cR(k)$, $f_\pb(M)$, and $\Omega_{\rm GW}(f)$ simultaneously. In the case of one perturbation, the PBHs with $f_\pb(M)\sim 1$ in the two small-mass windows at $10^{-17}~M_\odot$ or $10^{-13}~M_\odot$ can be achieved to compose all DM, and the relevant SIGWs are expected to be observed by the next-generation GW detectors, such as SKA, IPTA, LISA, and BBO, without touching the current constraint from aLIGO. As for the PBH of $30~M_\odot$, although its abundance is restricted to be merely $10^{-7}$ (more stringent than other constraints available in this mass window), it may still interpret the potential isotropic stochastic GW background from the NANOGrav 12.5-year dataset.

(3) In the parameter adjustments, with $\phi_0$ increasing, the peak of ${\cal P}_{\cal R}(k)$ moves to larger scales, the PBH mass $M$ increases, and the peak of $\Omega_{\rm GW}(f)$ moves to lower frequencies. Meanwhile, the most influential ingredient is the width $\sg$ of the perturbation. On the one hand, a larger $\sg$ corresponds to a longer duration of the USR stage and a larger PBH abundance, with $f_\pb(M)$ being exponentially dependent on $\sg$. On the other hand, the slope of the decreasing region of $\Omega_{\rm GW}(f)$ is also significantly affected by $\sg$. As $\sg$ decreases, the duration of the USR stage shortens, so the decreasing stage of $\Omega_{\rm GW}(f)$ becomes steeper. Because $\sg$ influences both $f_\pb(M)$ and $\Omega_{\rm GW}(f)$, to break the parameter degeneracy, the third parameter $A_0$ is indispensable in the model.

(4) In the case of two perturbations, the situations are similar. For the PBHs of $10^{-17}~M_\odot$ and $10^{-13}~M_\odot$, both of their abundances can be set to be $0.5$, so that $f_\pb\sim 1$ in total. For the PBHs of $10^{-13}~M_\odot$ and $30~M_\odot$, the former alone is able to compose all DM. On the contrary, the abundance of the latter is strictly restricted to be around $10^{-9}$, so its possibility as a candidate of DM is safely excluded, but even such low abundance is already enough to interpret the NANOGrav signal. The situation is similar for the PBHs of $10^{-17}~M_\odot$ and $30~M_\odot$.

(5) Because of the interference between the two perturbations, some features appear in Figs. \ref{fig:two1713}--\ref{fig:two1730}. For instance, there is an overlap between the decreasing region of $\cP_\cR(k)$ on larger scales and the increasing region of $\cP_\cR(k)$ on smaller scales. Consequently, there are distortions in $\cP_\cR(k)$ and $\Omega_{\rm GW}(f)$.

In conclusion, by appropriately constructing the antisymmetric perturbations on the background inflaton potential, we are able to achieve the PBHs with desirable abundances via the GLMS approximation of peak theory in the three typical mass window at $10^{-17}~M_\odot$, $10^{-13}~M_\odot$, and $30~M_\odot$, respectively. At the same time, the corresponding SIGWs are expected to be observed by the next-generation GW detectors, without spoiling the current constraint. Moreover, the potential isotropic stochastic GW background from the NANOGrav 12.5-year dataset may also be interpreted from the SIGW accompanying the PBH of $30~M_\odot$, albeit the relevant abundance is too small to explain DM. Altogether, our work is a phenomenological exploration of the inflaton potential with suitable features and will be helpful to further model building of cosmic inflation.

\acknowledgments

We are very grateful to Cui-Yuan Dai, Bing-Yu Su, and Hao-Ran Zhao for fruitful discussions. We also deeply thank the anonymous Referee for his valuable comments on the SIGW spectra. This work is supported by the Fundamental Research Funds for the Central Universities of China (No. N170504015).

\end{document}